\documentclass[twocolumn,superscriptaddress,preprintnumbers,amsmath,10pt,aps,prl,nobibnotes]{revtex4} 
\usepackage{amsmath}
\usepackage[ruled,vlined]{algorithm2e}
\usepackage{amsfonts}
\usepackage{bm}
\usepackage{xcolor}
\usepackage{verbatim}
\usepackage{graphicx}
\usepackage{siunitx}
\usepackage{upgreek}

\usepackage{soul}
\usepackage[size=scriptsize]{todonotes}

\begin{document}

\title{Space squeezing optics: performance limits and implementation\\ at microwave frequencies}

\author{Michal Mrnka}
\email{M.Mrnka@exeter.ac.uk}
\author{Euan Hendry}
\affiliation{School of Physics and Astronomy, University of Exeter, Exeter, EX4 4QL. UK.}
\author{Jaroslav Láčík}
\affiliation{Department of Radio Electronics, Brno University of Technology, 616 00 Brno, Czech Republic.}
\author{Rachel A. Lennon}
\affiliation{School of Physics and Astronomy, University of Exeter, Exeter, EX4 4QL. UK.}
\author{Lauren E. Barr}
\affiliation{School of Physics and Astronomy, University of Exeter, Exeter, EX4 4QL. UK.}
\author{Ian Hooper}
\affiliation{School of Physics and Astronomy, University of Exeter, Exeter, EX4 4QL. UK.}
\author{David B.~Phillips}
\affiliation{School of Physics and Astronomy, University of Exeter, Exeter, EX4 4QL. UK.}


\begin{abstract}

Optical systems often consist largely of empty space, as diffraction effects that occur through free-space propagation can be crucial to their function. Contracting these voids offers a path to the miniaturisation of a wide range of optical devices. Recently, a new optical element - coined a ‘spaceplate’ - has been proposed, that is capable of emulating the effects of diffraction over a specified propagation distance using a thinner non-local metamaterial [Nat.\ Commun.\ 12, 3512 (2021)]. The {\it compression factor} of such an element is given by the ratio of the length of free-space that is replaced to the thickness of the spaceplate itself. In this work we test a prototype spaceplate in the microwave spectral region (20-23\,GHz) - the first such demonstration designed to operate in ambient air. Our device consists of a Fabry-Pérot cavity formed from two reflective metasurfaces, with a compression factor that can be tuned by varying the size of perforations within each layer. Using a pair of directive horn antennas, we measure a space compression factor of up to $\sim$\,6 over an NA of 0.34 and fractional bandwidth of 6\,\%. We also investigate the fundamental trade-offs that exist between the compression factor, transmission efficiency, numerical aperture (NA) and bandwidth of this single resonator spaceplate design, and highlight that it can reach arbitrarily high compression factors by restricting its NA and bandwidth.

\end{abstract}

\maketitle

\noindent {\bf Introduction}\\
Free-space optical devices implicitly rely on the redistribution of energy that occurs when light diffracts through empty space. For example, lenses, gratings and prisms typically modify an incident wavefront at an interface (or pair of closely spaced interfaces). Yet the desired effect of this modification only becomes apparent once the optical field has propagated some distance beyond the interface, e.g.\ by focusing a beam or separating it into distinct diffraction orders. This requirement for free-space propagation places limits on the minimum operational volume of a wide range of optical elements and devices, such as cameras, microscopes, telescopes and spectrometers. The issue of size becomes even more prominent at longer wavelengths in so-called quasi-optical systems common to the terahertz \cite{stantchev2017compressed,stantchev2020real}, millimetre-wave \cite{Barr2021} and microwave domains~\cite{Goldsmith1998}. In this regime, free-space diffraction is fundamental to the operation of antennas \cite{Hum2014, Dahri2018, Guo2021, Imaz-Lueje2020} and beam waveguides \cite{Degenford1964}, and this can lead to very large optical systems \cite{Cappellin2018}. 
\begin{figure}[b]
\centering
\includegraphics[width=1\linewidth]{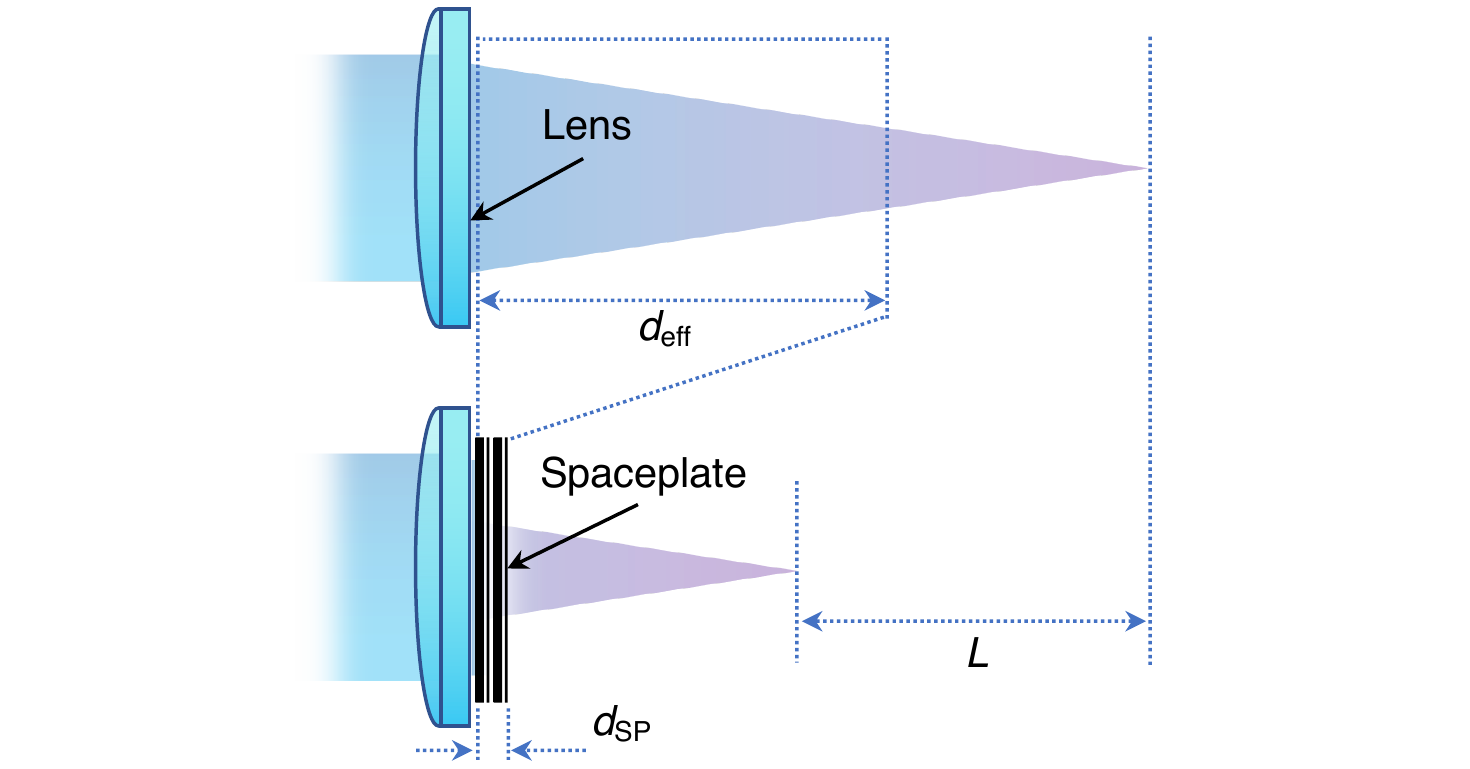}
\caption{The action of a spaceplate: replacement of a volume of free-space of length $d_{\mathrm{eff}}$ with an optical element of thickness $d_{\mathrm{SP}}$. In this example a spaceplate is shown in conjunction with a lens to move the focus closer without changing the numerical aperture. The amount of space contracted is $L = d_{\mathrm{eff}}-d_{\mathrm{SP}}$.}
\label{fig:spaceplate}
\end{figure}

Recently, Reshef et al.\ \cite{reshef2021} and Guo et al.\ \cite{Guo2020} introduced the intriguing new concept of a ‘spaceplate’ -- an optical element capable of mimicking the effects of free-space propagation. Crucially, a spaceplate's thickness ($d_{\mathrm{SP}}$) is thinner than the free-space distance it replaces ($d_{\mathrm{eff}}$), thus it can potentially be used to contract the volume of optical systems, as shown schematically in Fig.~\ref{fig:spaceplate}. The degree to which space is contracted is captured by the \textit{compression factor}, $\mathcal{C}$, given by the ratio of the emulated free-space propagation distance to the thickness of the spaceplate itself: $\mathcal{C} = d_{\mathrm{eff}}/d_{\mathrm{SP}}$. 

The effect of free-space propagation over a distance $d_\mathrm{eff}$ may be understood by decomposing an incident monochromatic optical field of wavelength $\lambda$ into its component plane-waves (i.e.\ spatial Fourier components). These plane-waves do not couple to one another during propagation, but each accumulates an angle-dependent phase shift of $\phi = k_z d_\mathrm{eff}$. Here, the wave-vector ${\bf k} = \left[k_x,k_y,k_z\right]$ describes the direction each plane-wave is travelling in Cartesian coordinates, and $k_z = k\cos\theta$, where wavenumber $k = 2\pi/\lambda = \left|{\bf k}\right|$, and $\theta$ is the polar angle of a plane-wave with respect to the optical axis. Therefore, in order to emulate free-space propagation, the action of a spaceplate must be `non-local'~\cite{elser2007nonlocal,capers2021designing}, i.e.\ it must independently act on the spatial Fourier components of the incident field, imparting an incident angle-dependent phase shift of
\begin{equation}\label{Eqn:phase}
    \phi_{\mathrm{SP}}\left(\theta\right) = k d_\mathrm{eff} \cos\theta.
\end{equation}

Designs fall into two main categories, which we term {\it stochastic} and {\it deterministic} spaceplates. Stochastic spaceplates, first introduced in ref.~\cite{reshef2021}, are non-local metamaterials consisting of a multi-layer stack of homogeneous and isotropic layers distributed along the optical axis. Structuring in 1D in this way ensures there is no coupling between plane-waves incident at different angles, as required. The parameters of individual layers -- the thicknesses and refractive indices -- can be algorithmically optimised to approximate a spaceplate with a target set of performance characteristics, within certain constraints (see Supplementary Information (SI) \S 5).

Deterministic spaceplates are founded on the understanding that certain families of structure readily act as space compressing optics. For instance, ref.~\cite{reshef2021} demonstrated that this is the case for a slab of material of lower refractive index than the surrounding medium. Guo et al.\ showed that the dispersion associated with a photonic crystal slab can be engineered so that the transmitted field components are imparted non-local phase shifts to mimic free-space propagation~\cite{Guo2020,long2022polarization}. Chen \& Monticone, meanwhile, highlighted that the angular dispersion in a Fabry-Pérot cavity operating slightly off resonance imparts close to the necessary angle-dependent phase shifts to transmitted light~\cite{chen2021}.

However, understanding the limitations on achievable spaceplate performance is an open problem. Of the different structures that have been theoretically shown to operate as spaceplates~\cite{reshef2021,Guo2020,chen2021,page2021designing,long2022polarization}, all exhibit some degree of trade-off between the key parameters defining performance: the \textit{compression factor}; the \textit{transmission efficiency} as a function of incident angle; the \textit{numerical aperture} (NA) and \textit{bandwidth} ($\delta\omega$) over which the spaceplate operates; and the total \textit{space contraction length} ($L$) -- see Fig.\ 1. For example, in all devices proposed so far, prioritising a high compression factor tends to reduce the achievable NA and bandwidth~\cite{page2021designing}. Furthermore, as the concept of space-compression optics is relatively new, the only experimental demonstration of a spaceplate to date relied on artificially increasing the refractive index of the ambient environment, and demonstrated a relatively modest compression factor of $\mathcal{C}\sim 1.2$ over the visible spectrum~\cite{reshef2021}. As such, it is not clear what maximum spatial compression is practically feasible in an air environment, where most envisioned applications lie.
\begin{figure}[t]
\centering
\includegraphics[width=1\linewidth]{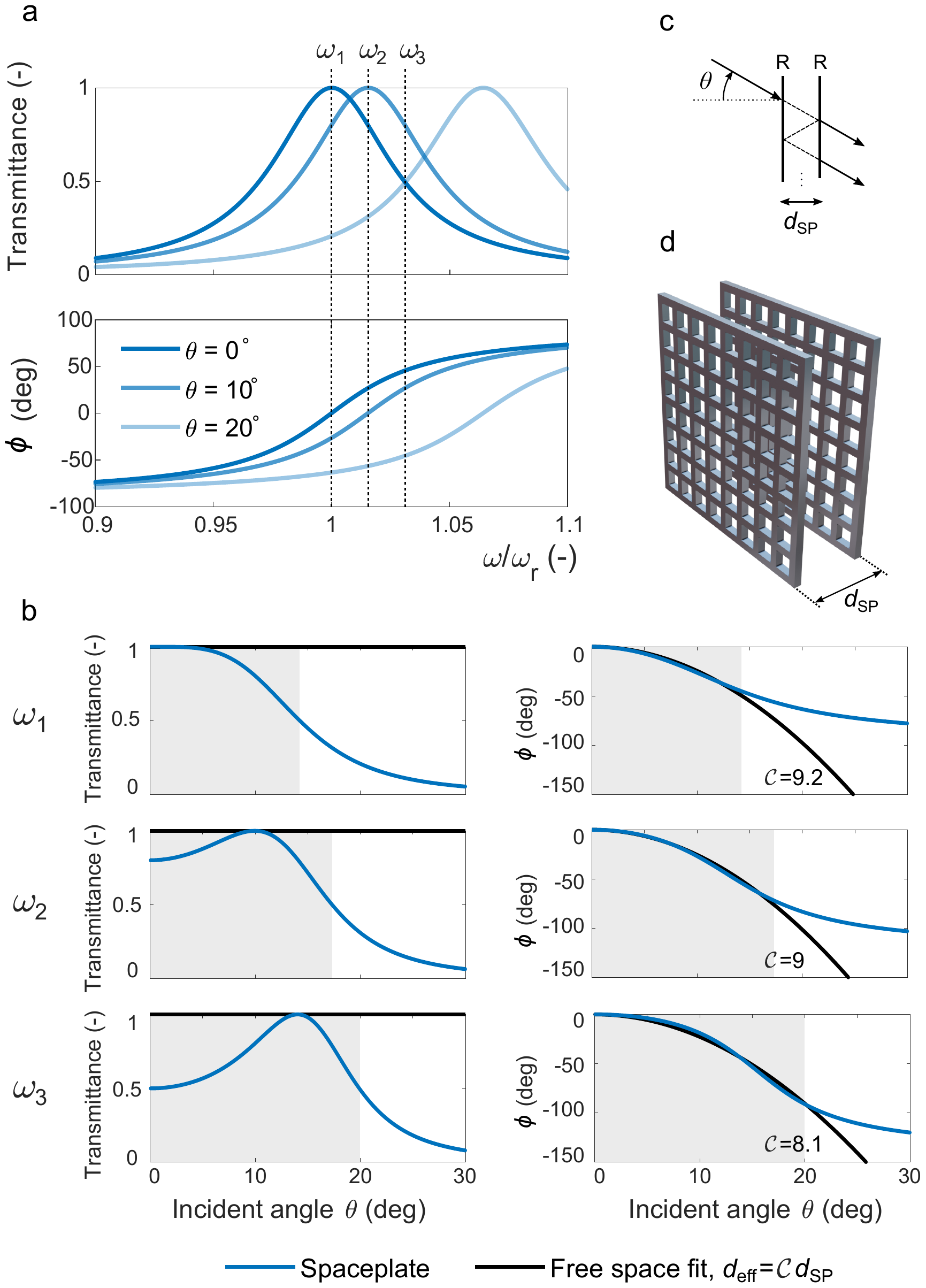}
\caption{The operation of a Fabry-Pérot cavity based spaceplate. In this example $R\sim0.8$ and $d_{\mathrm{SP}} = 7.1$~mm resulting in a $1^{\mathrm{st}}$ order resonance at $\omega_{\mathrm{r}}\sim1.3\mathrm{x}10^{11}$\,rad/s (21~GHz). {\bf(a)} The resonance line of a Fabry-Pérot cavity shifts approximately quadratically with incidence angle. {\bf(b)} The transmittance (left column) and phase (right column) as a function of incident angle, for three different angular frequencies near resonance ($\omega_1$, $\omega_2$, $\omega_3$). The maximum NA is achieved when the working angular frequency corresponds to $\omega_3 = \omega_r + \Delta\omega /2$, where $\omega_r$ and $\Delta\omega$ are the resonance frequency and the half power line width. While this maximizes the NA, it also introduces a 50\% reflective loss at normal incidence. Higher transmittance at normal incidence can be achieved if the angular frequency is closer to the resonance (e.g. $\omega_2$). The maximum transmittance and the lowest NA is achieved for $\omega_1 = \omega_r$. {\bf(c)} Schematic view of a general FP cavity formed by mirrors with equal reflectances $R$. {\bf(d)} Schematic view of the spaceplate proposed in this paper - the FP cavity is formed by two perforated conductive sheets acting as mirrors with tunable reflectance.}
\label{fig:spaceplate_operation}
\end{figure}

In this work, we design and experimentally test a prototype deterministic spaceplate operating in ambient air in the microwave region (20-23\,GHz). Our design consists of a two-layered resonator based on perforated conductive metasurfaces which form a Fabry-Pérot cavity. We study the performance limits of single resonator-based spaceplates and show that arbitrarily high compression factors may be reached by tuning the reflectance of the cavity mirrors. Experimentally, we demonstrate a spaceplate with a peak compression factor of $\mathcal{C}=6$. We discuss the advantages and drawbacks associated with our proof-of-principle prototype, and describe future improvements, which merge the concepts of deterministic and stochastic spaceplate design.\\

\noindent {\bf Space-compression using a Fabry-Pérot cavity}\\
Resonance features play a significant role in all ambient-air spaceplate deigns proposed so far. Therefore, following ref.~\cite{chen2021} we examine the potential of a single Fabry-Pérot cavity, operating slightly off-resonance, to act as a spaceplate. We consider a cavity composed of two semi-transparent mirrors with equal reflectance $R$, separated by a distance $d_{\mathrm{SP}}$, which corresponds to the thickness of the spaceplate (see Fig.~\ref{fig:spaceplate_operation}c). The complex transmission coefficient, as a function of plane-wave incident angle $\theta$ and angular frequency $\omega$, is the sum of the transmitted field components after successive passes around the cavity, which converges to
\begin{equation}
    t(\theta,\omega) = \frac{t_1t_2 \exp{(\mathrm{i} \beta)}}{1+r_1r_2 \exp{(2 \mathrm{i}} \beta)},
\end{equation}
where 
\begin{equation}
\beta = \frac{\omega}{c}d_{\mathrm{SP}} \cos{\theta}.
\end{equation}
Here $t_i$ and $r_i$ are the fraction of the amplitude of the incident wave transmitted or reflected at mirror $i$ respectively. For a single cavity in air, the Stokes relations connect transmission and reflection according to $r_1 = -r_2$, $r_1^2 = r_2^2 = r^2$ and $t_1t_2 = 1-r^2$, assuming absorption is negligible.

Within a limited angular range, the resonance frequency of a Fabry-Pérot cavity shifts approximately quadratically as a function of incidence angle -- see SI \S 1 for more detail. This results in a phase shift in the transmitted field that also depends quadratically on incident angle. Figure~\ref{fig:spaceplate_operation} shows an example of this effect. Figure~\ref{fig:spaceplate_operation}(a) shows the frequency of the resonance at three different incident angles. Figure~\ref{fig:spaceplate_operation}b shows the behaviour of the transmitted intensity $|t|^2$ and phase $\phi = \arg(t)$ as a function of incident angle, for three different frequencies located near to resonance. We can see that, as observed in ref.~\cite{chen2021}, within a finite numerical aperture (shaded in grey) the phase change as a function of incidence angle mimics that of free-space propagation over a distance $d_{\mathrm{eff}}$ that is larger than the spaceplate thickness $d_{\mathrm{SP}}$ (in this case yielding a compression factor of ${\mathcal{C} = d_{\mathrm{eff}}/d_{\mathrm{SP}}\sim8-9}$ that is weakly dependent upon frequency).
The transmitted intensity is $>$50\% over this angular range, although evidently transmission does vary as a function of incident angle. The largest operating NA occurs when the illumination frequency is slightly higher than the resonance frequency at normal incidence.
\begin{figure}[t!]
\centering
\includegraphics[width=1\linewidth]{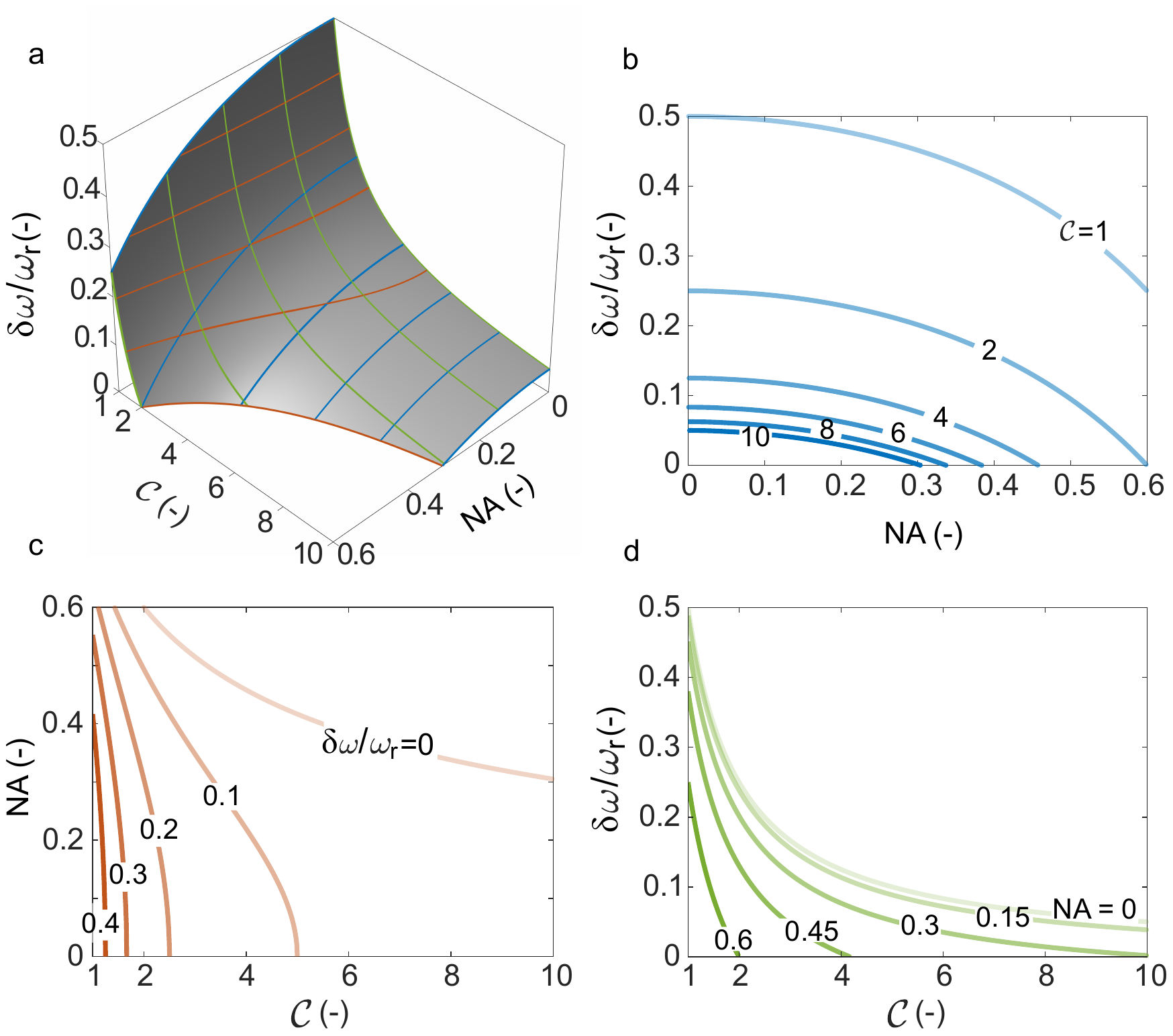}
\caption{Performance limits of Fabry-Pérot based spaceplates. ({\bf a}-{\bf d}) Allowed combinations of compression factor, NA and fractional bandwidth, constrained by ensuring the transmission efficiency $>50$\,\%.}
\label{fig:limits}
\end{figure}

Assuming high reflectance (so that $1-R<<1$), the compression factor $\mathcal{C}$ is related to the quality factor of the resonance ($Q$), and thus the reflectance $R$ of the cavity mirrors:
\begin{equation}\label{Eqn:compRef}
    \mathcal{C} = \frac{d_\mathrm{eff}}{d_{\mathrm{SP}}} \approx \frac{Q}{2\ell} \approx -\frac{\pi}{2\ln R},
\end{equation}
where $\ell$ is the order of the resonance (see SI \S 1.3 for derivation). Equation~\ref{Eqn:compRef} demonstrates that arbitrarily high compression factors may be reached by tuning the reflectance $R$ of the cavity mirrors alone, in a manner that is independent of $\ell$.  For example, a compression factor of $\mathcal{C}\sim$\,3200 can be reached by restricting the angular operating range to $1^\circ$. We note that this is a factor of $\sim$10 larger than the compression factor of $\mathcal{C}\sim$\,340 recently found by stochastic optimisation for an equivalent NA~\cite{page2021designing} -- a design also sharing a similar amplitude modulation function to a Fabry-Pérot based spaceplate. By further restricting the angular operating range of a Fabry-Pérot cavity to $0.5^\circ$, a compression factor of $\mathcal{C}\sim$\,13000 can be obtained. This behaviour can be encapsulated in a simple relation linking the NA to $\mathcal{C}$ and the desired operating bandwidth $\delta\omega$, found by considering the Q-factor of the resonance (see SI \S 1.4):
\begin{equation}\label{Eqn:NA}
    \mathrm{NA} \sim \left[1-\left[ 1+1/(2\mathcal{C}\ell)-\delta\omega/\omega_r \right]^{-2}\right]^{\frac{1}{2}},
\end{equation}
where $\omega_r$ is the angular frequency of the resonance. In deriving Eqn.~\ref{Eqn:NA}, we define the usable NA as the region over which the spaceplate transmits $>$50\% of incident power, while the desired bandwidth and compression factor must also satisfy $\delta\omega\leq\omega_r/(2\mathcal{C}\ell)$ to ensure the operational frequency range remains close to a resonance. In this analysis we have assumed that $R$ is nominally independent of frequency and incident angle.
Rearranging Eqn.~\ref{Eqn:NA} for the fractional bandwidth $\delta\omega/\omega_r$ yields:
\begin{equation}
    \delta\omega/\omega_r \sim 1 + (2\mathcal{C}\ell)^{-1} -
    (1-\mathrm{NA}^2)^{-\frac{1}{2}}.
    \label{Eqn:FBW}
\end{equation}
where here the choice of NA and compression factor must satisfy the condition
\begin{equation}\label{Eqn:BWconstraint}
    \mathrm{NA}^2\leq1-\left(1+(2\mathcal{C}\ell)^{-1}\right)^{-2}
\end{equation}
to ensure that the linewidth of the resonator is wide enough to accommodate the target compression factor over the specified NA.

Equations~\ref{Eqn:NA}~-~\ref{Eqn:BWconstraint} demonstrate the key trade-offs inherent in all spaceplate designs: higher compression factors are obtained at the expense of a reduced NA and bandwidth. Figure~\ref{fig:limits} illustrates these trade-offs, which will profoundly impact the applicability of high-compression spaceplates. SI  \S 1.4 also gives more detail.\\

\noindent {\bf Experimental demonstration of a spaceplate}

\noindent An important step is to explore the extent to which space-compression is readily achievable under experimental conditions. To investigate this, we have built a spaceplate based on two reflective metasurfaces which form a Fabry-Pérot cavity. Our prototype is designed to operate with a compression factor of up to $\mathcal{C}\sim 6$ over an NA of $\sim0.34$ in air (i.e.\ a maximum incident angle $\sim20^\circ$) and a frequency range of 20.8-22.1\,GHz.
We note that although the optical properties of Fabry-Pérot cavities are well-understood, here we experimentally study them from the novel perspective of space-compression.

The partially reflecting cavity mirrors are implemented with a simple metamaterial: a conductive sheet perforated with sub-wavelength sized square holes, as depicted in Fig.~\ref{fig:spaceplate_operation}d. These perforated layers must not couple the spatial Fourier components of transmitted radiation -- a constraint satisfied when the wavelength of radiation is larger than the period of the structure. Under this condition, the perforated metal behaves as homogeneous layer: a conductor with an effective plasma frequency considerably lower than the constituent metal~\cite{stone2011dispersion}. The effective, frequency-dependent permittivity of the layer is determined by the size and spacing of holes, allowing the creation of mirrors with well-controlled and near-arbitrary reflectivity. SI \S 2 discusses this metasurface based mirror design in more detail. Here we choose the geometry of the perforations to yield a reflectance of $R\sim$\,0.8. The cavity consists of two metasurfaces, each of area 0.3$\times$0.3\,m$^2$ separated to give an overall spaceplate thickness of $d_{\mathrm{SP}}\sim9.6$\,mm. In this case the 1$^{\mathrm{st}}$ order ($\ell = 1$) Fabry-Pérot resonance occurs at 21\,GHz (see SI \S 3.2). 

\begin{figure}[t]
\centering
\includegraphics[width=1\linewidth]{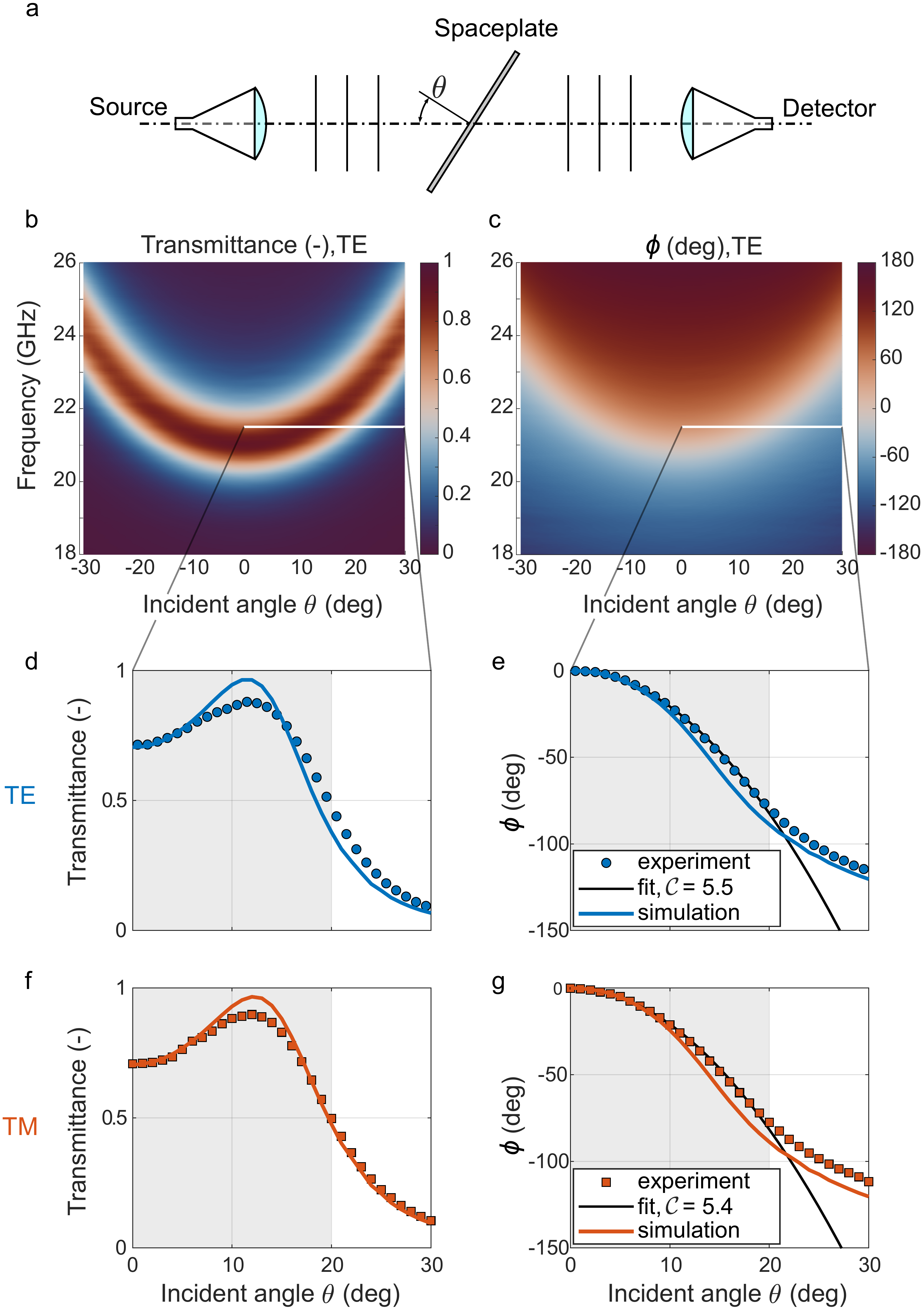}
\caption{Spaceplate dispersion measurement. ({\bf a}) Schematic of experimental setup. ({\bf b}) Transmittance and ({\bf c}) phase shift imparted by spaceplate as a function of incident angle and frequency (TE polarisation). ({\bf d}-{\bf e}) Cross-section cuts through the TE dispersion relations. (e) shows best fit free-space propagation phase function (black line), corresponding to a compression factor of $\mathcal{C} = 5.5$. ({\bf f}-{\bf g}) Cross-section cuts through the measured TM dispersion relations (see SI for full dispersion plots).}
\label{fig:transm_coeff_experimental}
\end{figure}

We first measure the dispersion of our prototype spaceplate, which is shown in Fig.~\ref{fig:transm_coeff_experimental}. The spaceplate is illuminated with a source approximating a plane-wave using a horn antenna, and the intensity and corresponding phase shift of the transmitted field is measured with a second horn antenna. Measurements are made as a function of polarisation, frequency, and incident angle -- adjusted by rotating the spaceplate with respect to the source and detector as shown in Fig.~\ref{fig:transm_coeff_experimental}(a). The data is collected using a vector network analyser (VNA), and normalised to the system response in the absence of the spaceplate. See SI \S4 for more details of this experiment.
\begin{figure*}[t]
\centering
\includegraphics[width=0.9\linewidth]{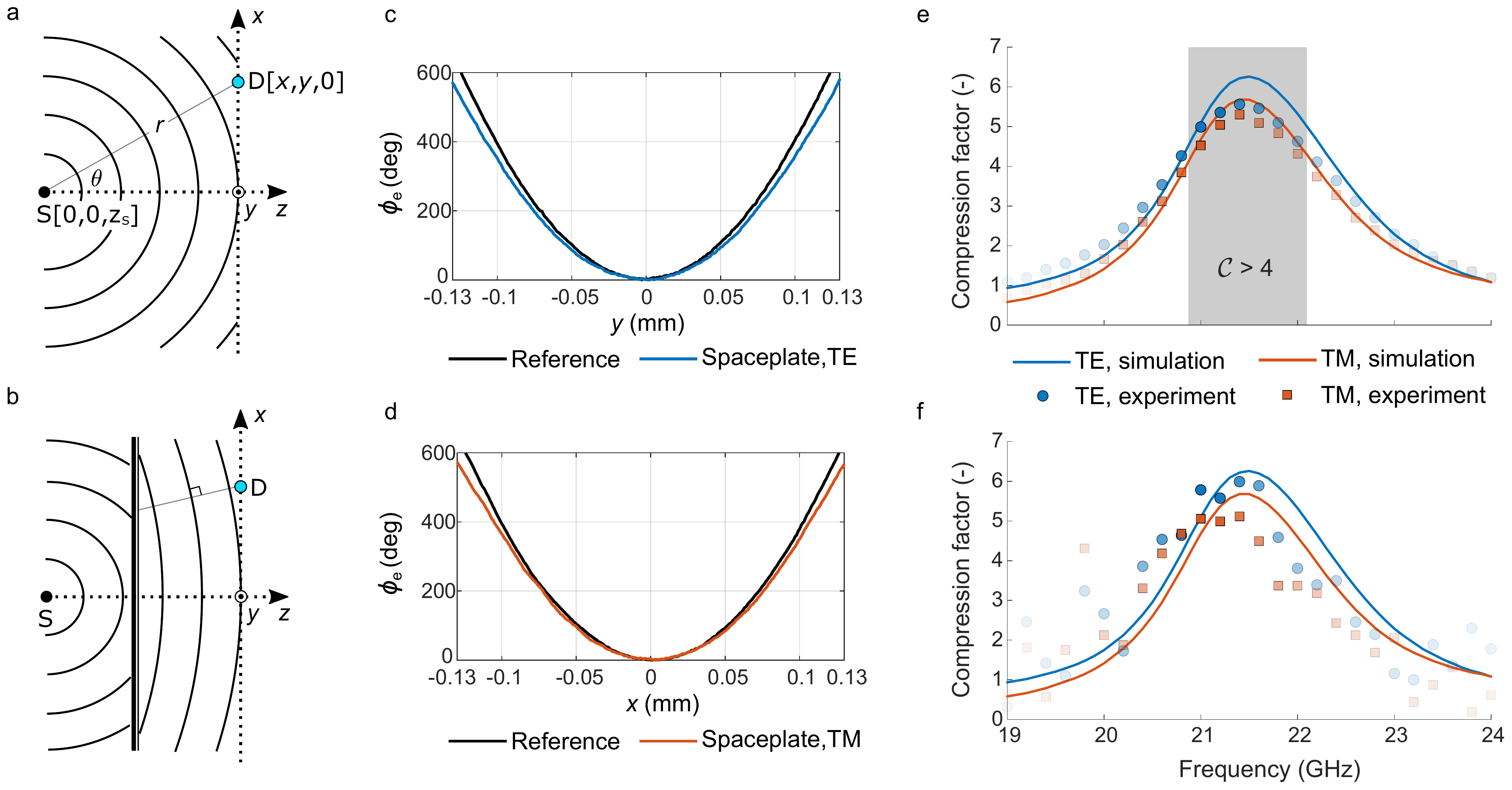}
\caption{({\bf a}) Schematic of field scan reference measurement. ({\bf b}) Schematic of spaceplate field scan measurement. ({\bf c}) The change in the wavefront curvature when the spaceplate is inserted into the measurement system (TE). ({\bf d}) Equivalent to (c) but for TM polarisation. ({\bf e}) Compression factor as a function of incident frequency, calculated from spaceplate dispersion measurement. ({\bf f}) Compression factor as a function of incident frequency, calculated from spaceplate field-scan measurement. The transparency of the points in (e) and (f) have been modulated in proportion to the transmitted intensity.
}
\label{fig:wavefronts_and_compression}
\end{figure*}

We observe a well-defined resonance band in the dispersion plots for both TE and TM incident radiation. Figures~\ref{fig:transm_coeff_experimental}(c-d)) show the TE case, see SI \S4 for the TM measurement. Figures~\ref{fig:transm_coeff_experimental}(d-e)) show angular cross-sections through the dispersion maps at a frequency of 21.5\,GHz. We see that the phase shift changes quadratically as a function of incident angle over the NA of the spaceplate (shaded in gray). To extract an estimate of the compression factor, we search for the effective propagation distance $d_{\mathrm{eff}}$ that has incident-angle-dependent phase-shifts that best fit our measurements. We find that the angular phase-shifts corresponding to a propagation distance of $d_{\mathrm{eff}}=53$\,mm closely fit to our experimental data (black line on Fig.~\ref{fig:transm_coeff_experimental}(e)), indicating a compression factor of ${\mathcal{C} =d_{\mathrm{eff}}/d_{\mathrm{SP}} = 5.5}$. Figures~\ref{fig:transm_coeff_experimental}(f-g) show the equivalent result for TM polarisation, yielding ${\mathcal{C} = 5.4}$.

We also verify the analytical design of our spaceplate with finite-element simulations incorporating the sub-wavelength scale features of the metasurface mirrors (performed in Ansys HFSS). These simulations also closely match our experimental results, shown as coloured lines in Figs.~\ref{fig:transm_coeff_experimental}(d-g). SI \S3 gives more detail of this numerical model.

Repeating this analysis for different illumination frequencies enables calculation of the compression factor $\mathcal{C}(f)$ as a function of illumination frequency $f$ -- signifying the bandwidth of our spaceplate. Figure~\ref{fig:wavefronts_and_compression}(e) shows the result of this calculation for both TE and TM polarisations. We observe reasonable agreement with our finite-element model: the slightly lower compression factors measured experimentally are due to inhomogeneous broadening of the resonance peak compared with that simulated (see SI \S3, Fig.~11.). A secondary effect may be caused by the illumination field only approximating a plane-wave, and in reality containing a small spread of wavevectors. This will also act to slightly reduce the measured Q-factor of the resonance.

When in use, a spaceplate must be able to separately address the spatial Fourier components of an incident wave without causing any coupling between them. While our dispersion measurements demonstrate the key characteristics of space compression when illuminated by individual plane-waves, it is also instructive to study our spaceplate's ability to operate on incident radiation containing many Fourier components simultaneously. Therefore, we next study the response of our spaceplate when illuminated by a diverging field, and spatially map the transmitted radiation. This enables direct measurement of the change in wavefront curvature when the spaceplate is introduced into the beam-path. In contrast to our dispersion measurements, this will be sensitive to any coupling between Fourier components due to unwanted scattering from defects or the edges of the spaceplate.

Our experiment is shown schematically in Fig.~\ref{fig:wavefronts_and_compression}(a-b). We position an antenna approximating a point-source $\sim150$\,mm behind the spaceplate, and measure the transmitted field along a horizontal line-scan, parallel to the plane of the spaceplate $\sim150$\,mm beyond it. The measured wavefronts, when illuminating with TE and TM polarisations, are shown in Figs.~\ref{fig:wavefronts_and_compression}(c-d), without (black-line) and with (coloured lines) the spaceplate present in the beam-path. As expected, we see the curvature of the wavefronts decreases when the spaceplate is inserted into the system, as the transmitted field emulates that of a field that has propagated further from the point source. The smooth variation in the phase of the field transmitted through the spaceplate indicates that scattering effects are minimal, as interference would appear as oscillations.

The frequency-dependent compression factor $\mathcal{C}(f)$ can also be calculated from this measured change in wavefront curvature. At each illumination frequency we find the apparent distance, $\delta z$, from the measurement plane to an ideal point source that has a phase curvature that best fits the measured phase curvature. We carry out this procedure on data with the spaceplate present ($\delta z_{\mathrm{SP}}$) and without the spaceplate in the system ($\delta z_{\mathrm{ref}}$). The length of space contracted is ${L(f) = \delta z_{\mathrm{SP}}(f) - \delta z_{\mathrm{ref}}(f)}$, and the compression factor is then inferred via ${\mathcal{C}(f) = (L(f)+d_{\mathrm{SP}})/d_{\mathrm{SP}}}$ (e.g.\ see Fig.~\ref{fig:spaceplate}). 

Figure~\ref{fig:wavefronts_and_compression}(f) shows the result of this calculation. Here the results follow the expected trend, although the data is noisier than the frequency-dependent compression factor calculated from the dispersion measurement, and a small systematic shift in the frequency of the peak position is observed in both polarisations.
There are two reasons why field-mapping measurements are more subject to error: they are obtained by scanning the position of the detection antenna, which can introduce minor systematic or random phase drifts due to the change in the configuration of the cable leading to the VNA during the scan. In addition, in this experiment the emitting antenna only approximates a point source, and so the apparent position of an ideal point source that best fits the data (also known as the `phase centre') can be subject to systematic errors that depend upon frequency. In both experiments we expect the error to be highest in the wings of the curves, where the intensity of the transmitted field is low and so any errors in phase measurements are magnified in these regions when the compression factor is calculated. To indicate this we have modulated the transparency of the data points in proportion to the transmitted intensity in the plots. In the region of high transmission, there is agreement between the compression ratios calculated using both measurement approaches, which indicates that our spaceplate is operating as designed. \\

\noindent{\bf Discussion}\\
All of the spaceplate designs proposed thus far \cite{reshef2021,Guo2020,chen2021,page2021designing,long2022polarization} exhibit trade-offs between the achievable compression factor, transmission efficiency, NA, bandwidth and total space contraction length. Here we have focussed on the simplest deterministic spaceplate design: a single Fabry-Pérot cavity, and highlighted that it can achieve a compression factor of nearly an order-of-magnitude higher than the largest quoted compression factor recently found by stochastic optimisation, over an equivalent NA~\cite{page2021designing}.

A single Fabry-Pérot cavity based spaceplate has two main drawbacks: there is a weak dependence of the compression factor on the incident polarisation (see Fig.~\ref{fig:wavefronts_and_compression}e), and transmitted intensity is modulated as a function of both frequency and angle (see Fig.~\ref{fig:spaceplate_operation}b). However, coupling together a series of Fabry-Pérot cavities provides opportunities to overcome these issues. For example, Chen \& Monticone theoretically showed that the transmission efficiency, along with the total space contraction length of a spaceplate, can be enhanced using coupled Fabry-Pérot cavities, whilst trading a modest reduction in compression factor~\cite{chen2021}.

Stochastic optimisation schemes also seem to generate spaceplate designs which are essentially coupled Fabry-Pérot cavities. For example, in order to overcome the limitations of a single Fabry-Pérot cavity, we have explored the use of a genetic algorithm to optimise the layer-spacing of a spaceplate consisting of up to 15 elements. SI \S 5 presents detail of our optimisation algorithm and results. We give the optimiser freedom to merge, and thus reduce, the number of layers when seeking an optimal solution. Using this method we find a locally optimal solution consisting of 3 Fabry-Pérot cavities separated by 2 optimised ($\sim\lambda/6$) coupling regions, which is very similar to the deterministic design presented by Chen \& Monticone~\cite{chen2021}. These coupled cavity spaceplate designs are able to suppress the polarisation dependence exhibited by a single Fabry-Pérot cavity, and generate a roughly constant transmission as a function of incident angle over the operating NA.

Taken together, the apparent dependence of all spaceplate designs on Fabry-Pérot resonance effects suggests that the trade-offs inherent in a single Fabry-Pérot cavity may be close to the fundamental and technical limits on spaceplate performance. We speculate that a Fabry-Pérot cavity may be understood as a basic building block of a spaceplate, in the same way as a simple lens is a basic building block of a multi-element (e.g.\ objective) lens. We envisage that the majority of future spaceplate designs will feature coupled Fabry-Pérot cavities, which will be honed for specific applications -- such as the optimisation of performance around three distinct colour channels for colour imaging -- in a similar manner to the way compound lenses are designed to suppress the chromatic and Seidel aberrations present in a single lens.

While in the process of updating our manuscript, a second study on the fundamental limits of spaceplates was released~\cite{shastri2022extent}. The bounds derived in this work are close to our analysis, while being marginally less restrictive, and a similar trade-off between compression factor, bandwidth and NA is shown. However, we note that while our exploration of spaceplate performance constrains the transmission efficiency to be greater than 50\%, the approach in ref.~\cite{shastri2022extent} does not constrain transmission efficiency. We would expect an improvement in performance beyond the limits given in our Eqns.~\ref{Eqn:NA} and~\ref{Eqn:FBW} might be possible by relaxing constraints on the acceptable level of transmission. As yet there is no concrete strategy to design spaceplates that can outperform Fabry-Pérot cavities -- it will be interesting to see if this becomes possible in the future.

It is also worth noting that the ability of a Fabry-Pérot cavity to non-locally modify the angular spectrum of electromagnetic waves is well-known to the antenna and microwave communities. Fabry-Pérot resonator antennas, first proposed by Trentini et al.\ in ref.~\cite{Trentini1956}, increase the directivity of an antenna by coupling it to a Fabry-Pérot resonator with considerably larger lateral dimensions~\cite{Liu2009}. It is also already understood that wavefronts emanating from the structure have travelled a longer path length than the thickness of the antenna, as theoretically derived by Burghignoli in ref.~\cite{Burghignoli2012}, which is consistent with the theoretical behaviour of a spaceplate. Although, we emphasise that in these earlier studies, the applications in mind were very different from the concept of space-compression.\\

\noindent{\bf Conclusions}\\
In summary, we have experimentally demonstrated a deterministically tunable space-squeezing optical element in the microwave spectral region. We observe a maximum space compression factor of $\sim$5.5 over an NA of 0.34 at 21.5\,GHz. The compression factor is higher than 4 in the frequency band 20.8-22.1\,GHz. We believe this is a significant step towards the introduction of the spaceplate concept into real-world quasi-optical systems. Our study hints that a Fabry-Pérot cavity may offer a close-to-optimal trade-off in capabilities, as encompassed by Eqn.~5, suggesting this type of simple spaceplace should be used as a benchmark for evaluating the performance of other designs.

\section*{Acknowledgments}
We thank Alex Powell for help with early parts of the project. The authors acknowledge financial support from the Engineering and Physical Sciences Research Council (EP/R004781/1, EP/S036466/1).
Initial measurements were supported by the Internal Grant Agency of Brno University of Technology, project no.\ FEKT-S-20-6526. DBP thanks the Royal Academy of Engineering and the European Research Council (804626) for financial support. \\

\section*{Disclosures}
 The authors declare no conflicts of interest.

\section*{Contributions}
MM devised the experiment and carried out the analytical and numerical modeling. IH built the spaceplate and conducted the experimental work. JL performed initial proof of concept measurements. LEB assisted with the simulations. RL assisted in the construction of the spaceplate. DBP and EH supervised the work. DBP, EH and MM wrote the manuscript. All authors edited the manuscript and contributed to the interpretation of the experimental data. \\

\bibliography{referencesMain}

\onecolumngrid
\setcounter{equation}{0}
\setcounter{figure}{0}
\vspace{50cm}
\noindent{\centering{\LARGE Supplementary Information}\par}
\vspace{5mm}

\noindent{This supplementary document provides further, in-depth information on the theory, various trade-offs and design of the Fabry-Pérot cavity based spaceplate. The experimental setup and methods used to characterise the spaceplate are explained here. In the last section we include information on the design and performance of a stochastic spaceplate consisting of three coupled resonators.}\\

\noindent{\large \bf \S 1 A Fabry-Pérot resonator as a spaceplate}\\
It has been shown that the angular dispersion in a Fabry-Pérot cavity can be used to design a spaceplate \cite{chen2021}. Here, we analyze such a solution from a general point of view and find the trade-off between the compression ratio ($\mathcal{C}$), numerical aperture (NA) and bandwidth ($\delta\omega$) of the spaceplate. In addition we show how the compression ratio can be obtained from the Q factor of the resonator or reflectances of its mirrors.

\begin{figure}[ht!]
\centering
\includegraphics[width=0.35\linewidth]{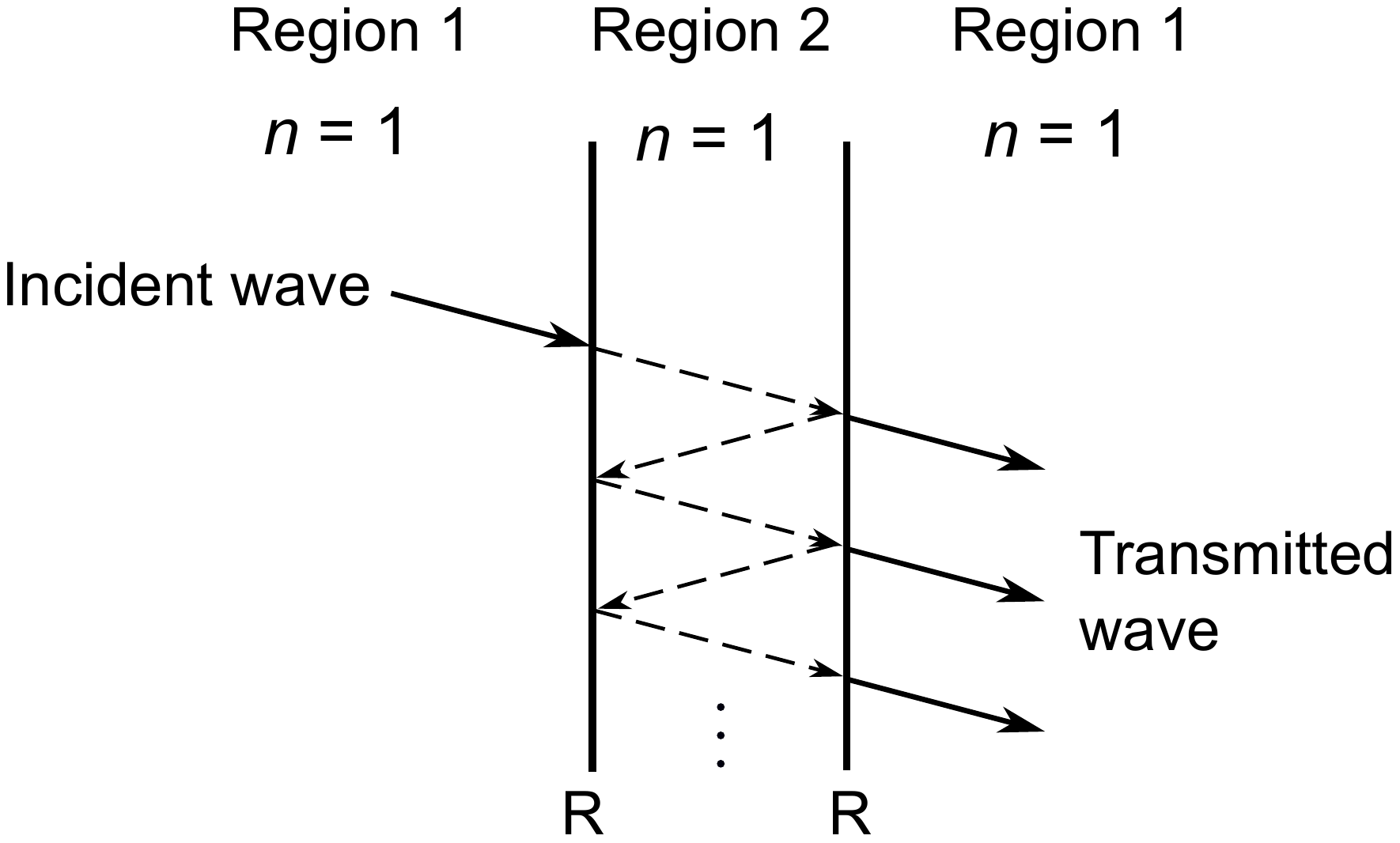}
\caption{Fabry-Pérot cavity formed by two mirrors with equal reflectances $R$.}
\label{fig:fp_cavity}
\end{figure}

The transmission coefficient of a Fabry-Pérot cavity made of two semitransparent mirrors with equal reflectance ($R_1 = R_2 = R$) separated by a distance $d_{\mathrm{SP}}$ is

\begin{equation}
    t(\theta, \omega) = \frac{t_{1} t_{2} \exp{(\mathrm{i} \beta)}}{1+r_{1} r_{2} \exp{(2 \mathrm{i}} \beta)},
    \label{eq:fabry_perot_trans}
\end{equation}
with 
\begin{equation}
\beta = \frac{2\pi}{\lambda} d_{\mathrm{SP}} \cos{\theta} = \frac{\omega}{c} d_{\mathrm{SP}} \cos{\theta},
\end{equation}
where $R = r^2$ and $r_1$, $r_2$, $t_1$ ,$t_2$ are the reflection and transmission coefficients of the mirrors (Stokes relations $r_1 = -r_2$, $t_1t_2 = 1-r^2$), $\lambda$ is the free space wavelength, $\theta$ is the incidence angle and $\omega$ is the angular frequency. We assume all the regions are free space (see Fig. \ref{fig:fp_cavity}).

\noindent{\bf 1.1 Even Fabry-Pérot resonances}\\
The interesting behavior for a single FP resonator spaceplate occurs near the transmission peaks where the conditions for constructive interference are met. The relative phase shift between two successive reflections defining the constructive interference is $2\beta =  2\pi \ell$, where $\ell$ is the order of the resonance ($\ell = 1, 2, 3...$). If we consider a cavity with thickness $d_\mathrm{SP} = \lambda_{r1}/2$, where $\lambda_{r1} = 2\pi\mathrm{c}/\omega_{r1}$ is the resonant wavelength of the first interference peak at normal incidence (i.e. $\theta = 0\,^{\circ}$), we can tie the resonance of the structure to the interference condition as follows:
\begin{equation}
     2\beta =  2\pi \ell \rightarrow 2\frac{2\pi}{\lambda_r(\ell, \theta)} d_{\mathrm{SP}} \cos{\theta} = 2\pi \ell \rightarrow  \frac{\omega_r(\ell, \theta)}{\mathrm{c}} \frac{2\pi\mathrm{c}}{\omega_{r1}} \cos(\theta)= 2\pi \ell .
    \label{eq:interference_condition}
\end{equation}
The dispersion equation relating the resonant frequency $\omega_r(\ell, \theta)$ corresponding to the angle $\theta$ and resonance order $\ell$ with respect to the normal incidence (and $\ell = 1$) resonance $\omega_{r1}$:
\begin{equation}
     \omega_r(\ell, \theta) = \frac{\ell}{\cos{\theta}} \omega_{r1}.
    \label{eq:multiresonance}
\end{equation}

\begin{figure}[ht!]
\centering
\includegraphics[width=0.8\linewidth]{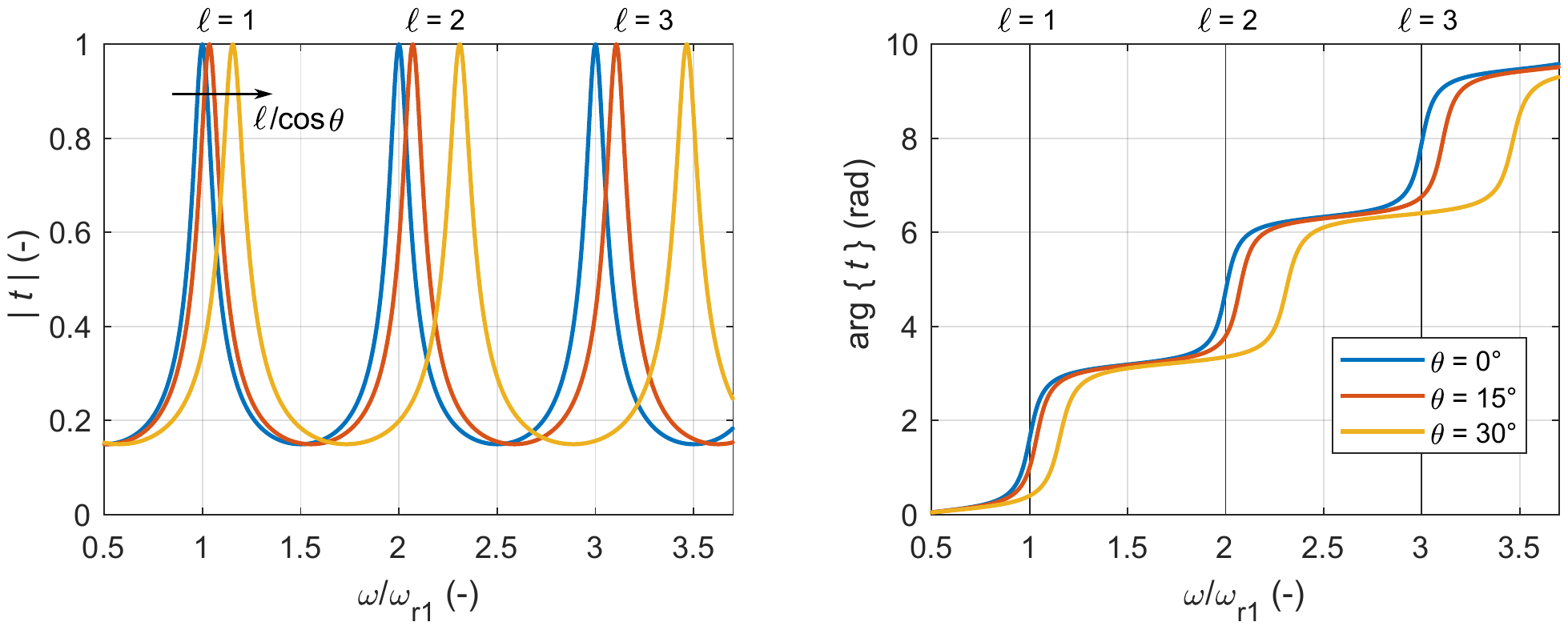}
\caption{First three even resonances and their shift with incidence angle - magnitude (left) and phase (right) of the transmission coefficient. The shift of $\ell^{th}$ order with angle $\theta$ is given by $\ell/\cos{\theta}$}
\label{fig:fp_resonances_freq}
\end{figure}

The resonant frequency of the FP cavity thus shifts with $1/\cos{\theta}$. However, for small angles in can be approximated by the first two terms of its Taylor expansion:

\begin{equation}
    1/\cos{\theta}\sim1+\frac{1}{2}\theta^2.
\end{equation}
Equation \ref{eq:multiresonance} tells us that the higher order (even) resonant modes ($\ell>1$) are more sensitive to the incidence angle by a factor $\ell$ (see Fig. \ref{fig:fp_resonances_freq}). This suggests that they provide a limited NA compared to the first one ($\ell = 1$), however as we show later, the compression ratio $\mathcal{C}$ remains the same. Effectively, this means that a resonator operating in its higher resonance can substitute a thicker slab of air but over a reduced angular range.

Now, we evaluate the phase of the transfer function of the spaceplate given by the dispersion according to eq. (\ref{eq:multiresonance}). To do this, we use a group delay (GD) concept described in section 1.2. We linearize the phase of the spaceplate about the resonance and the slope of the linear portion of the phase is considered to be the approximate GD.

\begin{figure}[ht!]
\centering
\includegraphics[width=0.5\linewidth]{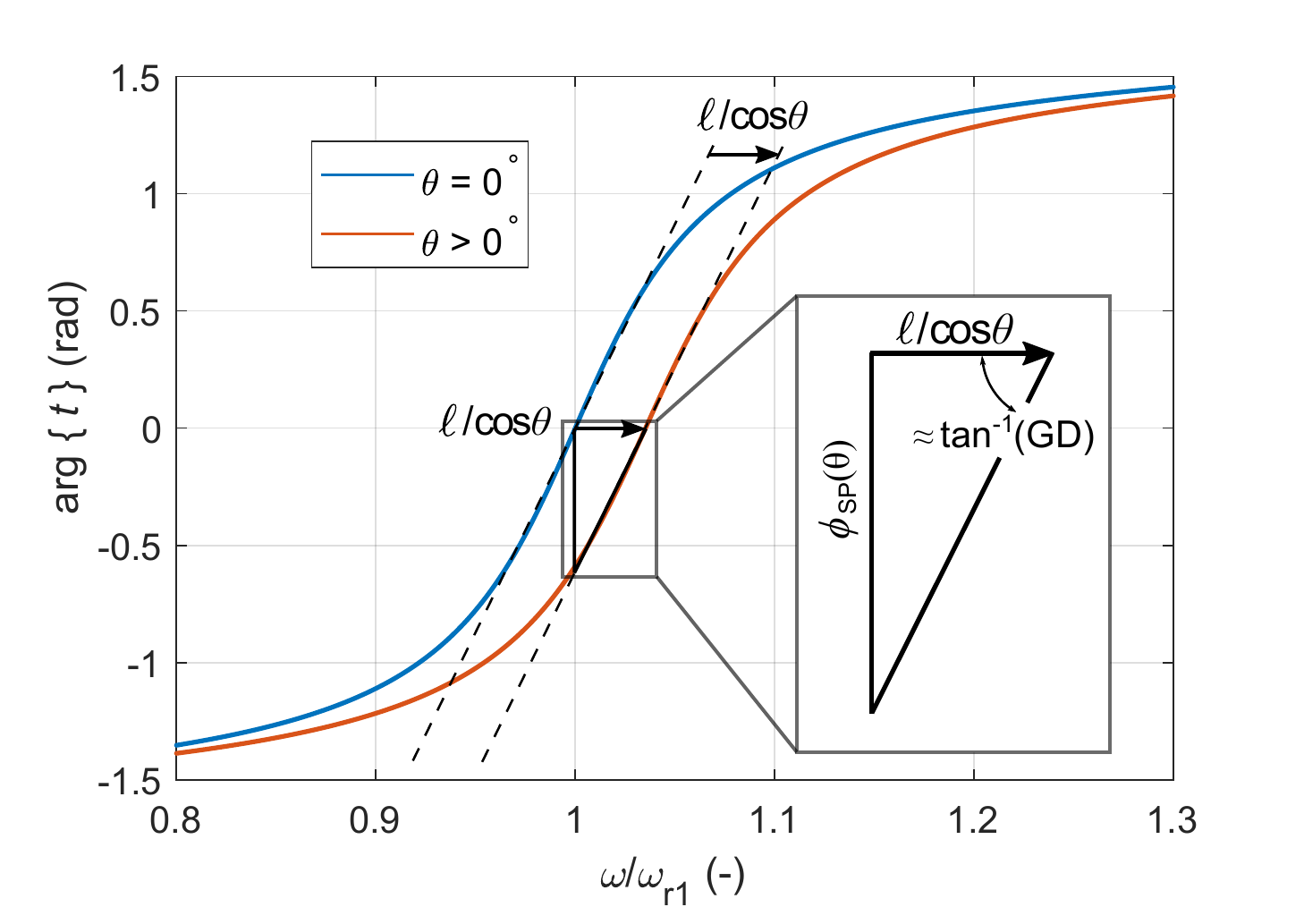}
\caption{Phase profile of the SP as a function of normalized angular frequency close to the 1$^{st}$ resonance. Blue curve represents normal incidence - the phase profile of the resonance is shifted by $\ell/\cos{\theta}$ (where $\ell = 1$) for incidence angle $\theta > 0\,^{\circ}$, plotted in red. The change in phase (arg {$\{t\}$}) related to the change of incidence angle $0\rightarrow \theta$ at operating frequency $\omega/\omega_{r1} = 1$ is $\phi_{\mathrm{SP}}(\theta)$.}
\label{fig:fp_resonator_phase_explained}
\end{figure}

Figure \ref{fig:fp_resonator_phase_explained} shows that the resonance of the spaceplate shifts with the incidence angle as $\ell/\cos{\theta}$. The phase $\phi_{\mathrm{SP}}(\theta)$ is the phase difference (over the length of the SP) between $\theta = 0\,^{\circ}$ and $\theta > 0\,^{\circ}$ angles. This phase can be found from the triangle in Fig. \ref{fig:fp_resonator_phase_explained} and is given as:

\begin{equation}
    \phi_{\mathrm{SP}}(\theta) = - \ell\frac{ GD}{\cos{\theta}}.
    \label{eq:sp_total_phase}
\end{equation}\\

\noindent{\bf 1.2 Compression factor of the spaceplate, $\mathcal{C}$, and the properties of the FP resonator}\\
\label{sec:fp_compr_factor}
The transmission curve of a resonator based spaceplate  is fully given by its quality factor (Q-factor) and the order of the resonance $\ell$. The compression factor $\mathcal{C}$ can be determined from the fundamental properties of the resonator (under certain assumptions) as follows.

\begin{figure}[ht]
\centering
\includegraphics[width=0.6\linewidth]{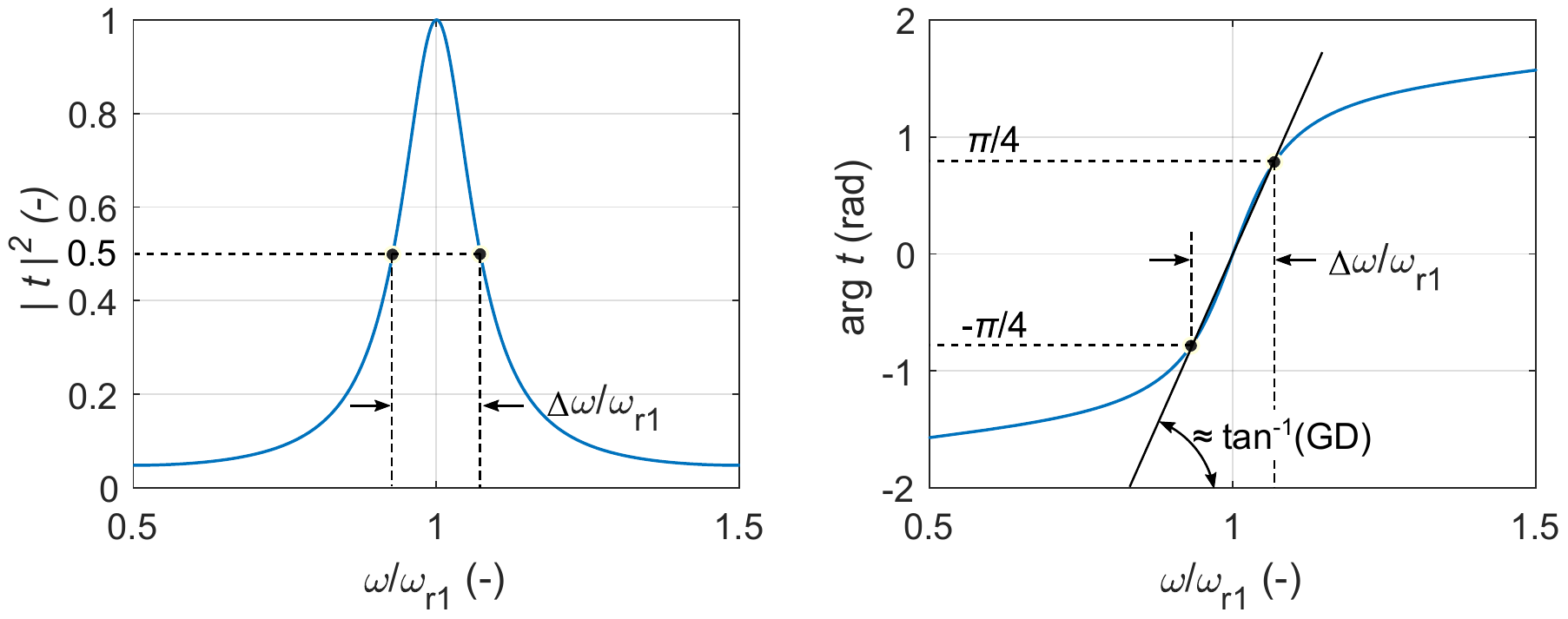}
\caption{Resonance line of a cavity with half wavelength thickness. The frequency bandwidth (full width at half maximum) is shown in the figures.}
\label{fig:fp_resonant_line}
\end{figure}

If we consider an FP cavity formed by 2 mirrors with identical reflectance $R$ and assuming:
\begin{itemize}
    \item  $R$ does not depend on the angle of incidence $\theta$, and
    \item the phase of the transmission coefficient is linear about the resonance (see Fig. \ref{fig:fp_resonant_line}),
\end{itemize}
then the slope of the linearised phase determines the group delay of the structure

\begin{equation}
    GD_{\mathrm{SP}} = \frac{\mathrm{d}\phi}{\mathrm{d}\omega} \approx \frac{\Delta \phi}{\Delta \omega} = \frac{\pi/2}{2\pi\Delta f} = \frac{1}{4\Delta f} = \frac{Q}{4f_r} = \frac{\pi Q}{2\omega_r},
\end{equation}
which must not change with the incidence angle and $\Delta f = \Delta \omega /2\pi$ is the line width of the resonator under normal incidence, $f_r = \omega_r/2\pi$ is the resonant frequency and the Q-factor is defined by $Q = f_r/\Delta f = \omega_r/\Delta \omega$. 

The group delay in free space of effective thickness $d_\mathrm{eff}$ which the spaceplate is to substitute is:
\begin{equation}
    GD_{\mathrm{FS}}= \frac{\mathrm{d}\phi}{\mathrm{d}\omega} = \frac{\mathrm{d}}{\mathrm{d}\omega}  \left( \frac{\omega}{\mathrm{c}} d_{\mathrm{eff}} \right) = \frac{d_{\mathrm{eff}}}{\mathrm{c}}.
\end{equation}
Thus, by assuming $GD_{SP} = GD_{FS}$ we obtain the equivalent thickness of free space, $d_{\mathrm{eff}}$, our resonator can substitute: $d_{\mathrm{eff}} = \mathrm{c}\cdot \pi Q / (2 \omega_r) = \lambda_r Q /4$ .

If we operate the spaceplate of thickness $d_{SP} = \ell \lambda_r/2$ at a frequency that matches the resonant frequency of the resonator $f = f_r$ the transmission for the normal incidence is 1 and it slowly tapers off with incidence angle. The compression factor $\mathcal{C}$ in this case is given as:
\begin{equation}
    \mathcal{C} = \frac{d_\mathrm{eff}}{d_\mathrm{SP}} = \frac{\frac{\lambda_r Q} {4}}{\ell \frac{\lambda_r}{2}} = \frac{Q}{2\ell}.
    \label{eq:compr_from_q}
\end{equation}

Assuming a high reflectance ($R$) of the mirrors that form the cavity (such as $1-R<<1$) we can tie the Q-factor to the reflectance as \cite{Renk2012}
\begin{equation}
    Q = - \frac{\pi \ell}{\ln R},
\end{equation}
which combined with eq. \ref{eq:compr_from_q} leads to the to the compression factor as a function of the reflectance of the mirrors:
\begin{equation}
    \mathcal{C} = -\frac{\pi}{2\ln R}.
    \label{eq:compr_refl}
\end{equation}
Under the assumptions given above, we can see that the compression factor $\mathcal{C}$ of the FP resonator spaceplate does not depend on the order of the resonance $\ell$. However, we should keep in mind that the higher resonances will offer a reduced NA (see eq. (\ref{eq:multiresonance})).\\

\noindent{\bf 1.3 Numerical aperture and bandwidth of the FP resonator spaceplate }\\
As shown below, there is a fundamental trade-off between the compression ratio, the numerical aperture of the spaceplate, and the frequency bandwidth it can operate over. 

\begin{figure}[ht]
\centering
\includegraphics[width=0.75\linewidth]{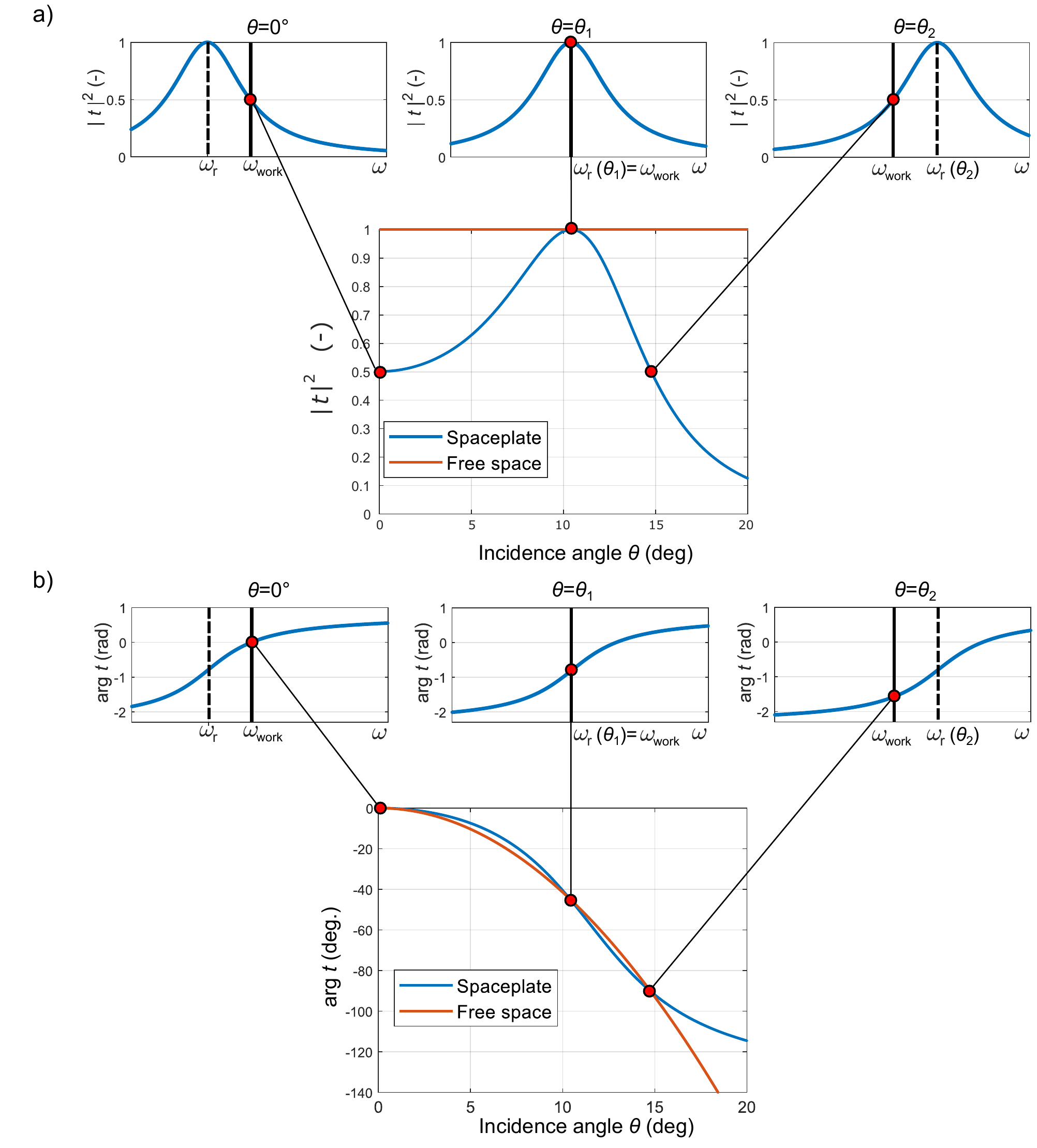}
\caption{Operation of the spaceplate explained using how the amplitude (a) and phase (b) of the transfer function depend on the resonance shift of the FP cavity with the incidence angle. This case corresponds to the maximum NA, where the transmittance at normal incidence is reduced to 0.5. The operating angular frequency in  is $\omega_{work} = \omega_r +\Delta \omega/2$, where $\omega_r$ is the resonant frequency at normal incidence. The important angles $\theta_1$ and $\theta_2$ are shown here.}
\label{fig:fp_explained_amplitude_phase}
\end{figure}

Starting with monochromatic waves, if we allow the transmittance of the spaceplate at normal incidence to be half of the maximum, the NA is maximized. The angles $\theta_1$ and $\theta_{2}$ describe the angles where the transmittance is increased to 1 (angle $\theta_1$) and where it drops to 0.5 again (angle $\theta_{2}$). In this case, the angle $\theta_{2}$ defines the NA = sin $\theta_{2}$. The operating angular frequency is  $\omega_\mathrm{work} = \omega_r + \Delta \omega/2$ (see Fig. \ref{fig:fp_explained_amplitude_phase}).

If we wish to have the transmittance at normal incidence equal to 1, the numerical aperture is given by the $\theta_1$ angle, NA = sin $\theta_1$ (operating frequency $\omega_\mathrm{work} = \omega_r$). 

The angles $\theta_1$ and $\theta_{2}$ are given as follows:

\begin{equation}
    \theta_1 = \cos^{-1}{\frac{\omega_r}{ \omega_r + \Delta \omega/2}} = \cos^{-1}\frac{1}{1 + 1/2Q} = \cos^{-1}\frac{1}{1 + 1/4\mathcal{C}\ell},
\end{equation}

\begin{equation}
    \theta_2 = \cos^{-1}{\frac{\omega_r}{\omega_r + \Delta \omega}} = \cos^{-1}\frac{1}{1 + 1/Q} = \cos^{-1}\frac{1}{1 + 1/2\mathcal{C}\ell}.
\end{equation}
The numerical aperture for the case with transmittance at normal incidence equal to 0.5 can be written using the trigonometric identity $\sin(\cos^{-1}({x})) = \sqrt{1-x^2}$
\begin{equation}
    NA = \sin\theta_2  = \sqrt{1-\left( \frac{1}{1+1/Q} \right)^2 } = \sqrt{1-\left( \frac{1}{1+1/2\mathcal{C}\ell} \right)^2 },
    \label{eq:na_compression}
\end{equation}
and it can be easily modified for the case with unity transmittance  at 0\,$^{\circ}$, $NA = \sin{\theta_1}$.

Now, we turn our attention from monochromatic waves and introduce a signal with a certain angular frequency bandwidth $\delta \omega = \omega_{\mathrm{max}} - \omega_{\mathrm{min}}$ and due to the nature of the spaceplate we must assume $\delta \omega < \Delta \omega$ (if we do not want to compromise on the performance even more than shown in Fig. \ref{fig:fp_explained_amplitude_phase}). This effectively means that the maximum acceptable frequency shift of the resonance, which is $\Delta \omega$ for monochromatic waves will be reduced to $\Delta \omega - \delta \omega$. Using the dispersion equation (eq. (\ref{eq:multiresonance})) the maximum angle $\theta_2$ is

\begin{equation}
    \theta_2 = \cos^{-1}{\frac{\omega_r}{\omega_r + \Delta \omega - \delta \omega}} = \cos^{-1}\frac{1}{1 + 1/Q - \delta\omega/\omega_r} = \cos^{-1}\frac{1}{1 + 1/2\mathcal{C}\ell - \delta\omega/\omega_r},
\end{equation}
and the NA of the spaceplate with given guaranteed frequency bandwidth $\delta\omega$ is
\begin{equation}
    NA = \sin\theta_2  = \sqrt{1-\left( \frac{1}{1+1/Q-\delta\omega/\omega_r} \right)^2 } = \sqrt{1-\left( \frac{1}{1+1/2\mathcal{C}\ell-\delta\omega/\omega_r} \right)^2 }.
    \label{fig:na_bandwidth_compression}
\end{equation}

The equation \ref{fig:na_bandwidth_compression} can be rearranged for the approximate fractional bandwidth:
\begin{equation}
    \delta \omega / \omega_r = 1 + \frac{1}{2\mathcal{C}\ell} -\frac{1}{\sqrt{-NA^2+1}},
    \label{fig:na_bandwidth_compression_2}
\end{equation}
where $ 0 \leq \delta\omega/\omega_r \leq 1/2\mathcal{C}\ell$ interval bounds the feasible solutions. It enforces that the bandwidth is a positive quantity $ 0 \leq \delta\omega/\omega_r$ and that we work sufficiently close to the resonance $ \delta\omega/\omega_r \leq 1/2\mathcal{C}\ell$ to cap the maximum allowed transmission loss at 0.5. We call the $\delta \omega / \omega_r$ quantity an "approximate" fractional bandwidth as the $\omega_r$ frequency is not at the centre of the band $\delta\omega$.
The equations are very accurate for high Q cavities, but still work well for lower Q cases ($R\approx 0.5$).\\

\begin{figure}[h]
\centering
\includegraphics[width=0.68\linewidth]{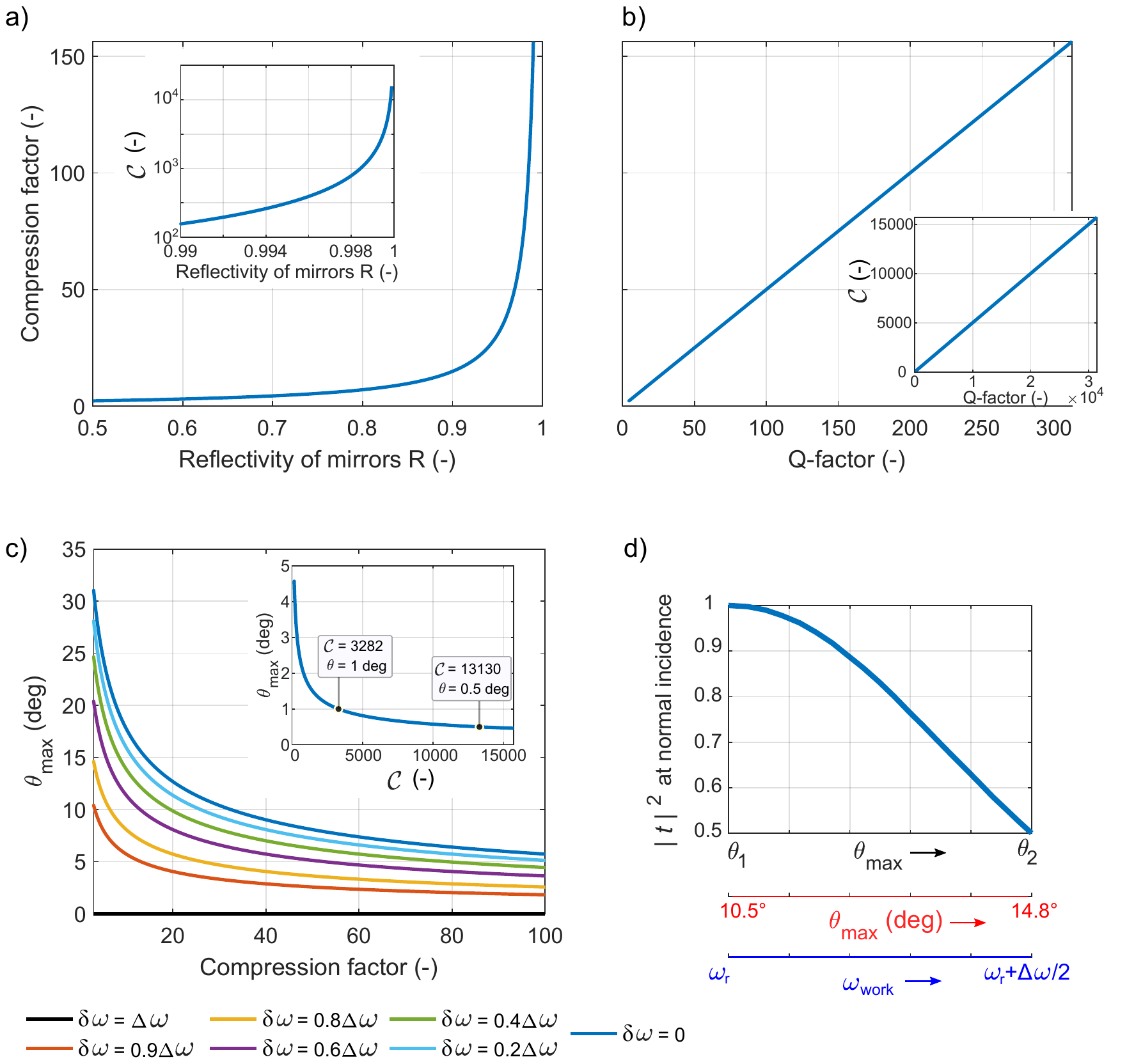}
\caption{Main trade-offs of a FP resonator spaceplate (with $\ell = 1$). $\delta\omega$ represents the available bandwidth and $\Delta\omega$ is the linewidth of the resonator}
\label{fig:tradeoffs_1}
\end{figure}

\noindent{\bf 1.4 Interesting trade-offs}\\
Equations (\ref{eq:compr_from_q}), (\ref{eq:compr_refl}), (\ref{eq:na_compression}) and (\ref{fig:na_bandwidth_compression}) are useful for studying the performance trade-offs and the limits of FP resonator spaceplates. To do so, we  sweep the reflectance of the mirrors $R$ in the range 0.5 to 0.99 and plot the compression factor as a function of $R$ (see Fig. \ref{fig:tradeoffs_1}a) and as a function of the Q-factor of the cavity (see Fig. \ref{fig:tradeoffs_1}b). The insets  show the compression factor $\mathcal{C}$ for even higher reflectances ranging from 0.99 up to 0.9999. In limiting case, for  $Q \rightarrow \infty$ (also $R \rightarrow \infty$), the compression ratio will tend to infinity (see eq. (\ref{eq:compr_from_q})) as the NA will be approaching 0 (eq. (\ref{eq:na_compression})):
\begin{equation}
    \lim_{Q \to +\infty} \mathcal{C} = \infty \\
    \lim_{Q \to +\infty} NA = 0 
\end{equation}

An important trade-off for a resonant spaceplate is that between the numerical aperture (or maximum incidence angle $\theta_{\mathrm{max}}$), the compression factor and the bandwidth ($\delta \omega$). In Fig. \ref{fig:tradeoffs_1}c we can see the maximum acceptance angle (transmittance = 0.5 at normal incidence) as a function of the compression factor with the bandwidth as a parameter. The bandwith is expressed as a fraction of the spectral linewidth of the resonator $\Delta \omega$. The blue curve ($\delta \omega$ = 0) represents a monochromatic wave and thus sets the upper bound on achievable numerical aperture. As the bandwidth increases the NA goes down. The maximum theoretical bandwidth (i.e. $\delta \omega = \Delta \omega$) does not allow for any shift of the resonance and thus results in NA = 0. The inset in Fig. \ref{fig:tradeoffs_1}c shows that a compression factor of $\mathcal{C}=3282$ can be achieved with FP resonator spaceplate within NA corresponding to $\theta_{max} =1^{\circ}$, and $\mathcal{C}=13130$ within $\theta_{\mathrm{max}} =0.5^{\circ}$.

It has been mentioned that the spaceplate trades off its numerical aperture for the transmittance at normal incidence (Fig. \ref{fig:fp_explained_amplitude_phase}a). Figure 2 of the main paper shows that this depends on the selection of the operating frequency $\omega_{\mathrm{work}}$. If we choose to operate at resonance frequency $\omega_{\mathrm{work}} = \omega_r$ we achieve maximum transmittance $|t|^2$=1 at normal incidence, which tapers off to $|t|^2$ = 0.5 at incidence angle $\theta_1$. On the other hand, if we pick $\omega_{\mathrm{work}} = \omega_r+\Delta \omega/2$, the transmittance at normal incidence is $|t|^2$ = 0.5, increases to $|t|^2$=1 (at $\theta_1$), and again drops to $|t|^2$ = 0.5 at incidence angle $\theta_2$. In Fig. \ref{fig:tradeoffs_1}d we can see the drop in normal incidence transmittance as a function of maximum incidence angle $\theta_{\mathrm{max}}$. The red axis demonstrates a real world scenario of a resonant spaceplate based on mirrors with $R=0.9$ and a compression factor $\mathcal{C}=14.9$ - here we can see that by allowing $|t|^2$ = 0.5 at normal incidence we can increase the NA angle from $\theta_{\mathrm{max}} =10.5\,^\circ$ to $\theta_{\mathrm{max}} =14.8\,^\circ$. Finally, the blue axis in Fig. \ref{fig:tradeoffs_1}d relates the range of maximum angles to the selection of operating frequency $\omega_{\mathrm{work}}$.

\begin{figure}[ht!]
\centering
\includegraphics[width=0.35\linewidth]{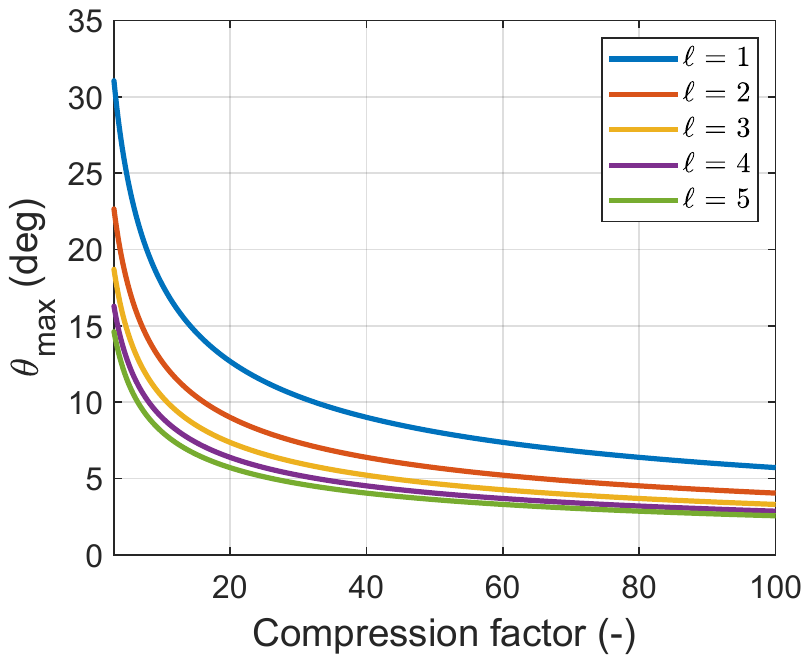}
\caption{Numerical aperture as a function of compression ratio for the 5 lowest order resonances, assuming monochromatic operation. The drop in NA with increasing $\ell$ is obvious.}
\label{fig:tradeoffs_2}
\end{figure}

Even though the resonance order $\ell$ does not influence the maximum compression ratio of a spaceplate, it has a negative effect on its achievable numerical aperture (see eq. (\ref{eq:multiresonance})). In Fig. \ref{fig:tradeoffs_2} we examine this effect for $\ell =1$ to $\ell = 5$. 
The main benefit of operating the spaceplate at a higher order resonance is the increase in effective thickness $d_\mathrm{eff}$ since the length of the resonator is increased $d_\mathrm{SP} = \ell \cdot \lambda/2$ and the compression factor remains the same, $d_\mathrm{eff} = \mathcal{C}\cdot d_\mathrm{SP}$.
The second benefit is the mechanical simplicity with low weight and costs. The spaceplate consists of only three components - two dichroic mirrors separated by free space.\\

\noindent{\large \bf \S 2 Metamaterial properties of perforated metal layers}\\
The electromagnetic interaction between light and the metal is driven by the free electrons of the metal, which give rise to a negative permittivity for frequencies below their plasma frequencies. It has been shown that one can mimic the negative permittivity of a metal near its plasma frequency by structuring the surface of a (near) perfect conductor by introducing periodic arrays of subwavelength holes. Such a structured layer of highly conducting material can support surface plasmon-like surface modes at frequencies well below the plasma frequency of the conductor \cite{HO_Ref1,HO_Ref2, HO_Ref3,stone2011dispersion}, and also have the property of partial reflectance, similar to that of thin (less than the skin depth) metal films \cite{HO_Ref5}.

In this paper, we use highly conducting sheets with two-dimensional arrays of subwavelength, open holes to mimic the partially reflecting mirrors of a Fabry Pérot cavity. It is important that perforated layers are non diffracting, which is the case when the wavelength of radiation is larger than the period of the structure. Under this condition, the perforated metal behaves as homogeneous layer, and propagating plane waves are expected to be de-coupled \cite{HO_Ref3}, a necessary condition for a space plate. The effective, frequency dependent permittivity of such a layer is determined by the size and spacing of holes, allowing the creation of mirrors with well-controlled and near-arbitrary reflectivity. When both the perforation holes and the period of the array is significantly less than the wavelength of incident radiation, one expects a large negative effective permittivity at low frequencies, similar to that seen for homogeneous metals at frequencies well below the plasma frequency \cite{HO_Ref3}.\\

\noindent{\large \bf \S 3 Design of a resonant spaceplate}\\
A Fabry-Pérot resonator spaceplate can be built similarly as shown in section \S 5 of SI as an air (free-space) cavity surrounded by high index dielectric sheets. Indeed, this approach was proposed in \cite{chen2021}, where the authors suggested using thin sheets of material with $\varepsilon_\mathrm{r} = 15$ operating near the quarter wavelength resonance to maximize the reflectance of the sheets. Here, we use metallic hole arrays (described in the previous section) made of thin copper sheets to design the mirrors of the FP cavity. This approach has several practical advantages such as its low cost, reduced thickness and, most importantly, the tunability. By adjusting the filling ratio one can design a dichroic mirror with arbitrary reflectance. This in turn gives direct and continuous control over the Q-factor of the cavity, thus defining the compression ratio of the spaceplate. The analytical model of the hole arrays that we use to determine their reflectance was described in the previous section. The model allows us to include the angle dependent reflectance of the mirrors in eq. (\ref{eq:fabry_perot_trans}). We verify the analytical design of the spaceplate with numerical simulations in a commercial finite element software, Ansys HFSS.\\

\begin{figure}[ht]
\centering
\includegraphics[width=1\linewidth]{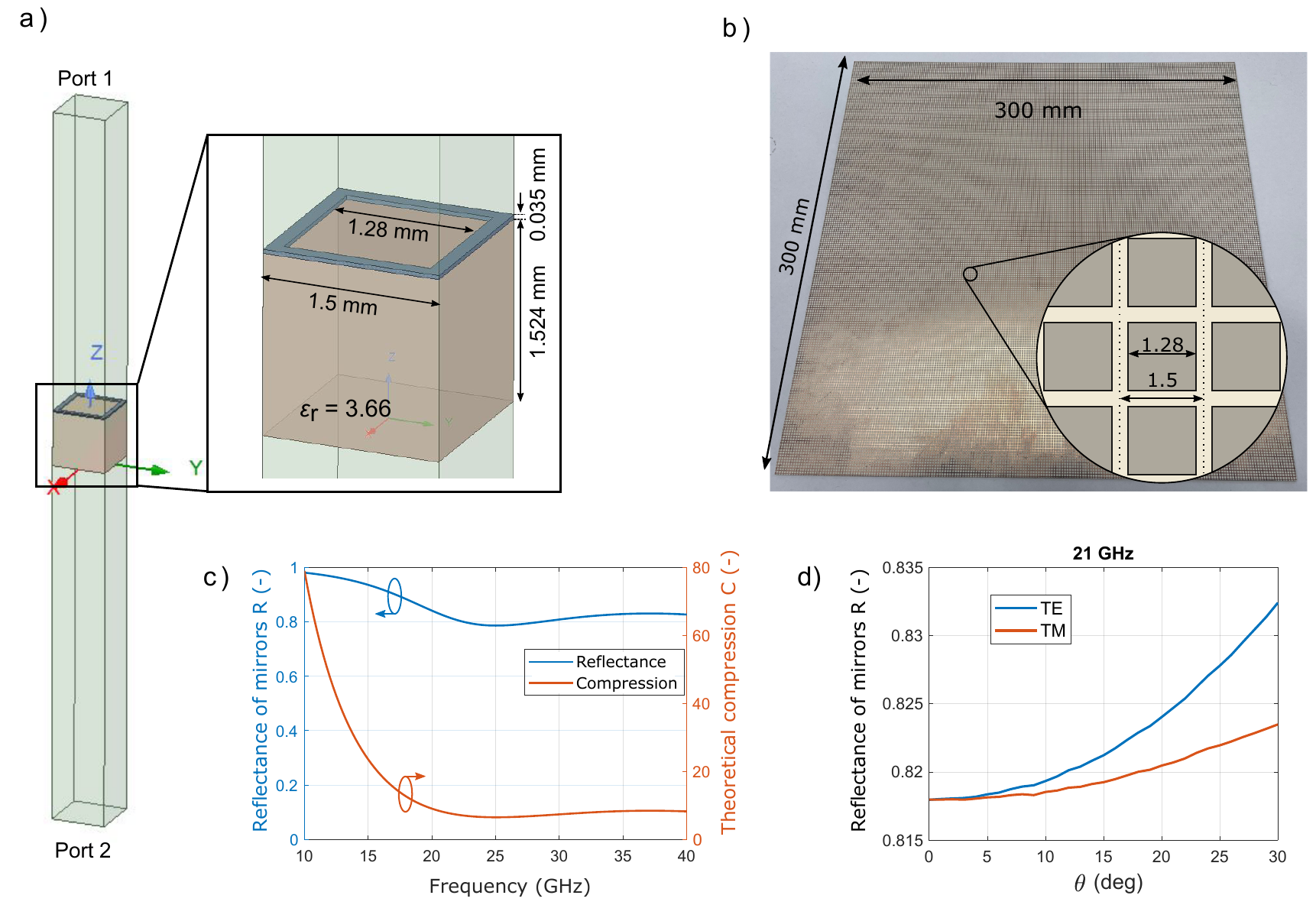}
\caption{ a) Unit cell of the periodic mirror, Floquet ports are applied to the top and bottom area of the unit cell, all the other faces are set up as periodic boundaries. The inset shows a detail of the unit cell with dimensions. b) The photograph of the dichroic mirror used to build the spaceplate with a detail of the square unit cell c) Reflectance of the mirror and theoretically achievable compression ratio of a spaceplate based on the mirror as a function of frequency. d) Reflectance of the mirrors as a function of incidence angle for the TE and TM polarisation at 21~GHz. }
\label{fig:dichroic_mirror_unit_cell}
\end{figure}

\noindent{\bf 3.1 Analysis and design of dichroic mirrors}\\
\label{sec:numerical_dichroic}
The mirrors are implemented by etching a periodic square hole motive on a copper cladding (35 $\upmu$m thick) of a dielectric microwave substrate Rogers R4350B with thickness 1.524~mm, dielectric constant $\varepsilon_\mathrm{r} = 3.66$ and loss tangent tan$\delta$ = 0.0037 (10~GHz). To achieve a reasonable trade-off among bandwidth, compression factor and NA we target a reflectance of the mirrors close to 0.8 ($\mathcal{C} = 7$). The final dimensions of the fabricated hole array sheets can be found in Fig. \ref{fig:dichroic_mirror_unit_cell}a.

We numerically analyze the hole array as a periodic structure with a unit cell shown in Fig. \ref{fig:dichroic_mirror_unit_cell}a using Ansys HFSS. First, we sweep the frequency in the range 10-40~GHz  and observe the reflectance (see Fig. \ref{fig:dichroic_mirror_unit_cell}c) of one sheet. From these values, using eq. (\ref{eq:compr_refl}), we can predict the compression ratio of a single FP resonator spaceplate made of two sheets, as a function of frequency. 

At operating frequency 21 GHz and normal incidence, the reflectance of the mirrors is about 0.82 which corresponds to the maximum theoretical compression factor of $C = 7.9$. Due to the final thickness of our dichroic mirrors deposited on dielectric substrates, the achievable compression is to be appreciably lower than the theoretical limit.

Unfortunately, the reflectance of the mirrors changes with the incidence angle in a way that differs for the two polarisation states (TE and TM). We plot this in Fig.~\ref{fig:dichroic_mirror_unit_cell}d, where we sweep the incidence angle in the range of $\theta = 0$ to $30$ deg. It is obvious that the discrepancy in reflectance between the two polarisations increases with increasing incidence angle - as we show later this directly influences the performance of the SP operating with the two polarisations by slighlty increasing the theoretical compression factor for TE waves while reducing it for the TM waves (see Fig.~5 of the main paper).\\

\noindent{\bf 3.2 Analysis and design of the experimental spaceplate}\\
The experimental spaceplate is formed from a pair of dichroic mirrors described in the previous section. The lateral size of the spaceplate is 300x300~mm$^2$ The mirror sheets are kept apart at constant distance by 6.56~mm thick spacers resulting in resonant frequency at normal incidence of 21~GHz.

\begin{figure}[h!]
\centering
\includegraphics[width=0.7\linewidth]{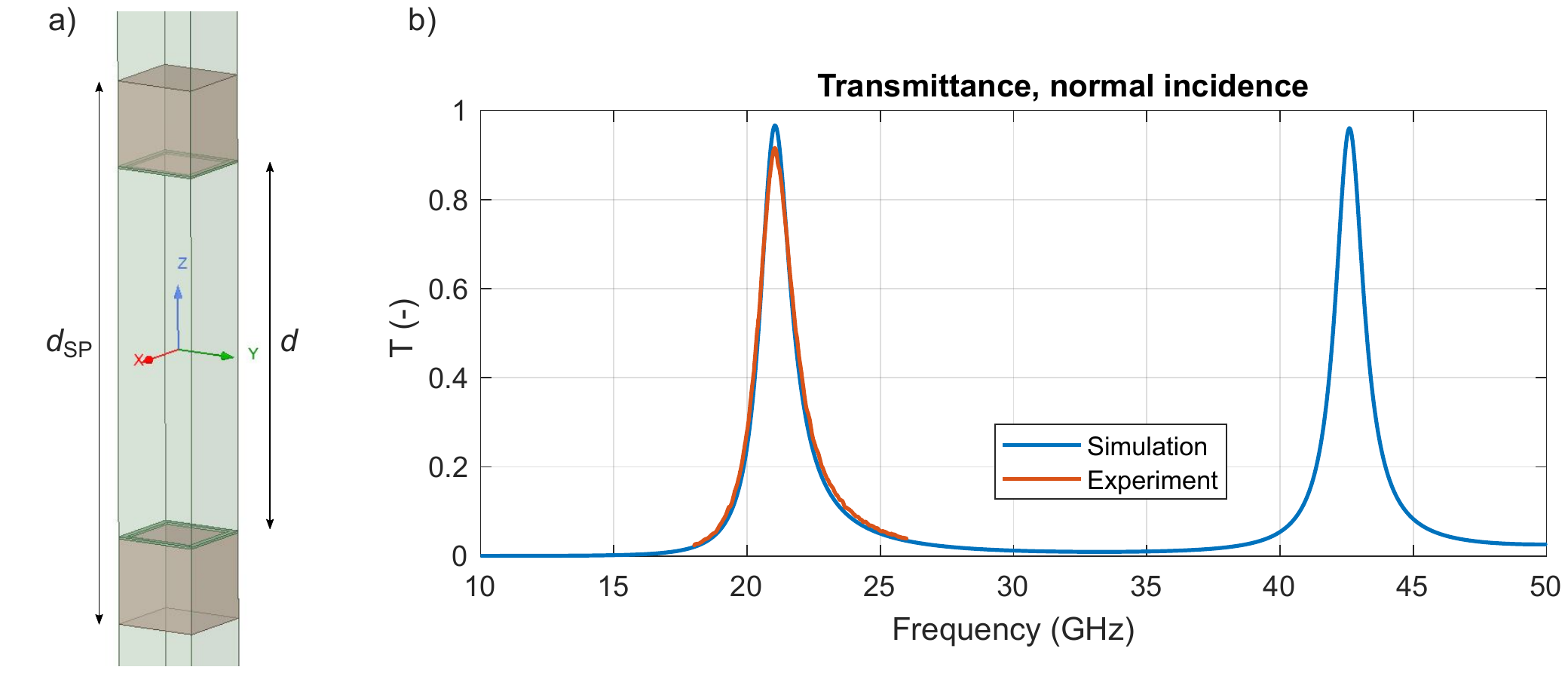}
\caption{ a) Unit cell of the resonant spaceplate. The distance between the mirrors $d$ = 6.56~mm and the thickness of the spaceplate $d_\mathrm{SP}$=9.618~mm b) Two lowest simulated resonance lines of the FP resonator spaceplate. Measured data for the $\ell$ = 1 line is included for comparison.}
\label{fig:s21_res_line_sim_meas}
\end{figure}

The structure is again analysed as a periodic structure with a unit cell given in Fig.~\ref{fig:s21_res_line_sim_meas}. The simulated spectral line is compared to the measured one in Fig.~\ref{fig:s21_res_line_sim_meas}. The experiment is described in the following section.

The simulated transmittance and the phase of the transmission coefficient  are shown in Fig.~\ref{fig:simulation_measurement_dispersion} at five frequency points 21, 21.2, 21.4, 21.6, 21.8\,GHz with $f_\mathrm{r} = 21$\,GHz. Both polarisations are shown in the figures together with a free-space fit. As a result of the unequal reflectances of the mirrors for the two polarisations, the phase of TE polarisation is slightly steeper than the TM polarisation. This corresponds to a slightly higher compression ratio and a smaller NA for the TE polarisation compared to the TM case. For convenience the measurement results are also included in these plots. The measurement setup and processing are described in detail in the following section.\\

\noindent{\large \bf \S 4 Experimental setup and results}\\
\label{sec:experiment}

Here, we describe the two experiments we performed to validate the performance of the spaceplate. First, a dispersion measurement where we measured a plane wave response of the spaceplate under varying angle and angular frequency. Second, a field mapping, where we directly measured the effect the spaceplate has on the radius of curvature of wavefronts radiated by a point source-like antenna.\\

\noindent{\bf 4.1 Dispersion measurement}\\
The normalised transmissivity, and corresponding phase change upon transmission, of 18 to 26 GHz radiation through the spaceplate were measured as a function of the angle of incidence using a pair of Flann 810 series lens horn antennas, which have a 150 mm diameter and produce a beam with a nominal 3dB beamwidth of 5.7 and 6.6 degrees in the E- and H-planes respectively with a mid-band gain of 29.7 dBi. These were placed 1.2 m apart, with a Thorlabs HDR50/M computer controlled rotation stage placed midway between them. ABS-ASF-12 partially absorbing foam layers from ABS-Technics were placed in front of each antenna in order to reduce the influence of standing waves resulting from reflections from the front faces of the antennas (and the sample when in place). The antennas were connected to an Anritsu Vectorstar MS4647B Vector Network Analyser, and the magnitude and phase of the transmitted signal between the horn antennas was measured as a function of incident angle with the spaceplate placed upon the rotation stage. These measurements were subsequently normalised to data obtained with no sample in place, with the normalised magnitude data squared to give the transmitted intensity, and the normalised phase data  giving the phase change upon transmission through the thickness of the spaceplate with respect to a slab of air with thickness equal to the spaceplate. The phase in Fig.~12, 13 and Fig.~4 of the main paper is thus calibrated to show the transmission phase of the spaceplate compared to zero thickness of air \cite{Friedsam1997} - the same definition as we use in our analytical and numerical models.   \\
\begin{figure}[h]
\centering
\includegraphics[width=1\linewidth]{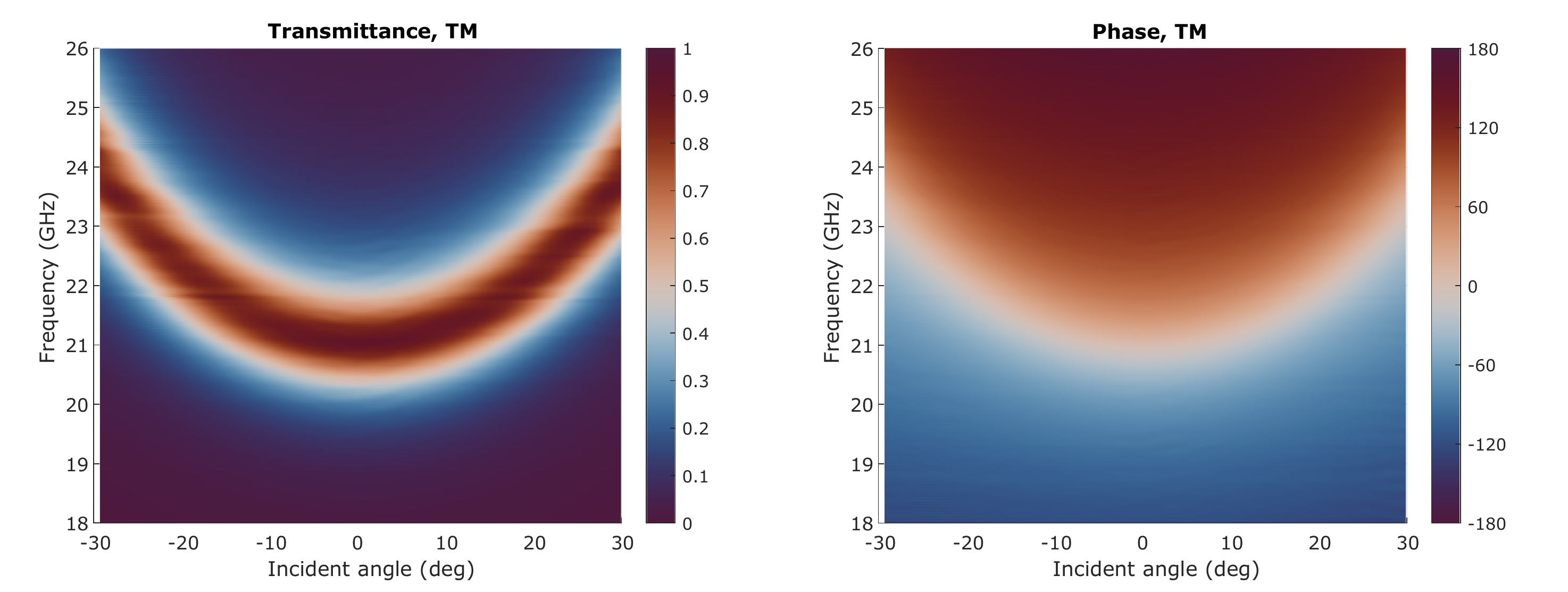}
\caption{ Dispersion plot for TM polarisation. TE polarisaton can be found in Fig.~3 of the main paper.}
\label{fig:dispersion_TM}
\end{figure}

\begin{figure}[p]
\centering
\includegraphics[width=1\linewidth]{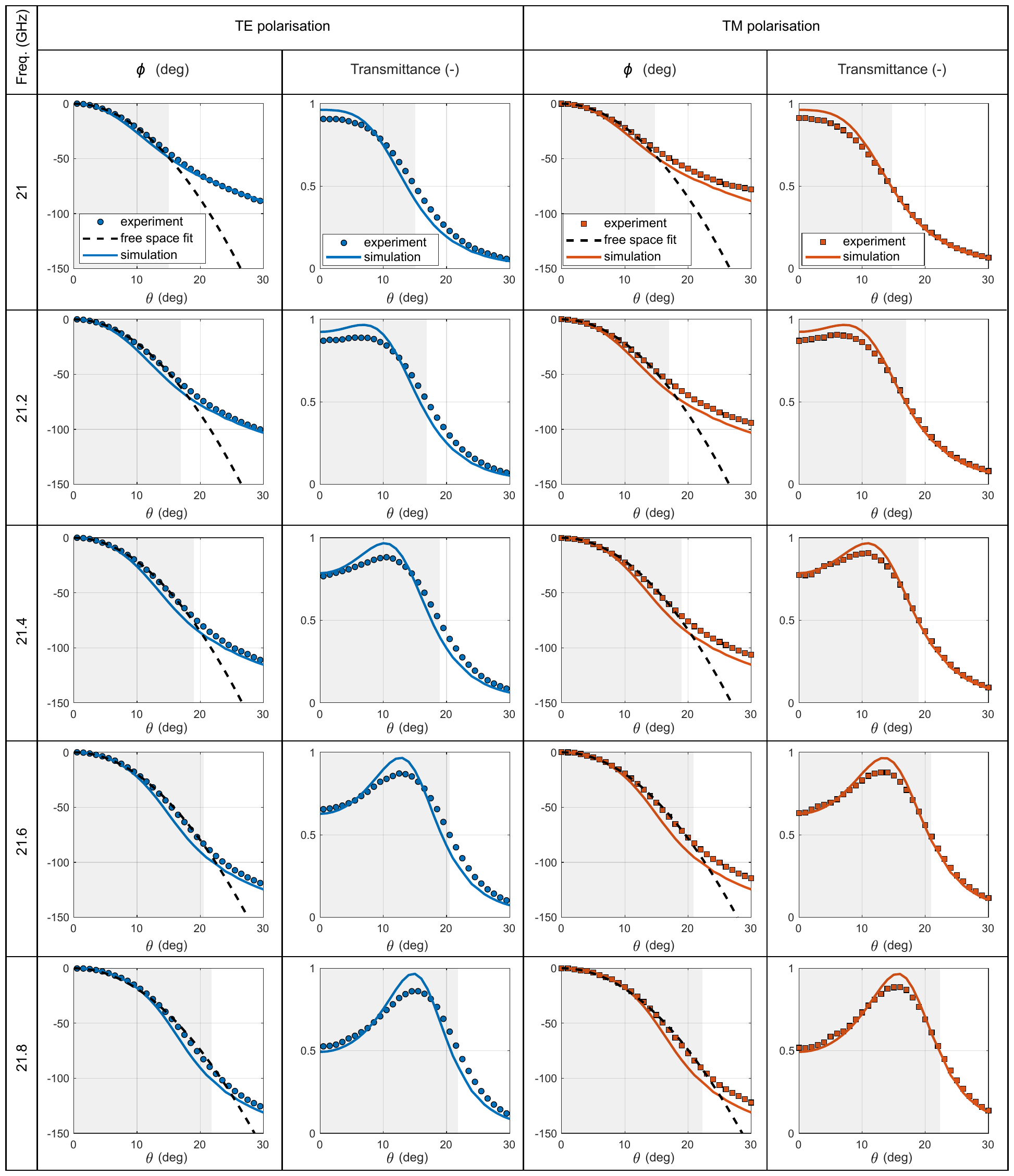}
\caption{ Transmission phase $\phi$ and transmittance as a function of plane wave incidence angle $\theta$ and frequency  for TE and TM polarisation. The plots compare HFSS simulation results (solid lines) with dispersion measurement results (markers) and also show a free space fit to the experimental data (dashed lines). The gray box represents the NA of the spaceplate - it is slightly smaller for the TE polarisation due to higher reflectance of the mirrors}
\label{fig:simulation_measurement_dispersion}
\end{figure}

\noindent{\bf 4.2 Spatial mapping of transmitted fields}\\

To directly characterise the wavefronts, a Narda 638 standard gain horn antenna (3dB beam width of between 34 and 23 degrees across the 18 to 26~GHz band) was connected to one of the ports of an Anritsu MS4644A vector network analyser (VNA), and placed 150~mm from the front face of the spaceplate such that the radiated field impinged upon the centre of the spaceplate at (and around) normal incidence. A stripped coaxial antenna with 4~mm of protruding central conductor was connected to the 2nd port of the VNA, and mounted on a computer controlled linear translation stage at a distance of 150~mm from the back face of the spaceplate such that the protruding end of the antenna could be translated through the mid-point of the beam. The overall length of the stripped coaxial antenna and mount was sufficient that the translation stage was outside of the beam area, with microwave absorber being distributed around the volume of the beam to ensure that the measurement ensemble was minimally perturbing to the beam. The magnitude $|t|$ and phase $\phi_e$ of the transmission between the emitting and detecting antennas was measured across a 300~mm line through the beam centre for both TE and TM polarisations with and without the spaceplate in place. This enabled a direct measure of the influence of the spaceplate on the wavefronts of the beam to be made (see Fig.~\ref{fig:measured_phase_and_magnitude}).\\

\begin{figure}[ht!]
\centering
\includegraphics[width=0.8\linewidth]{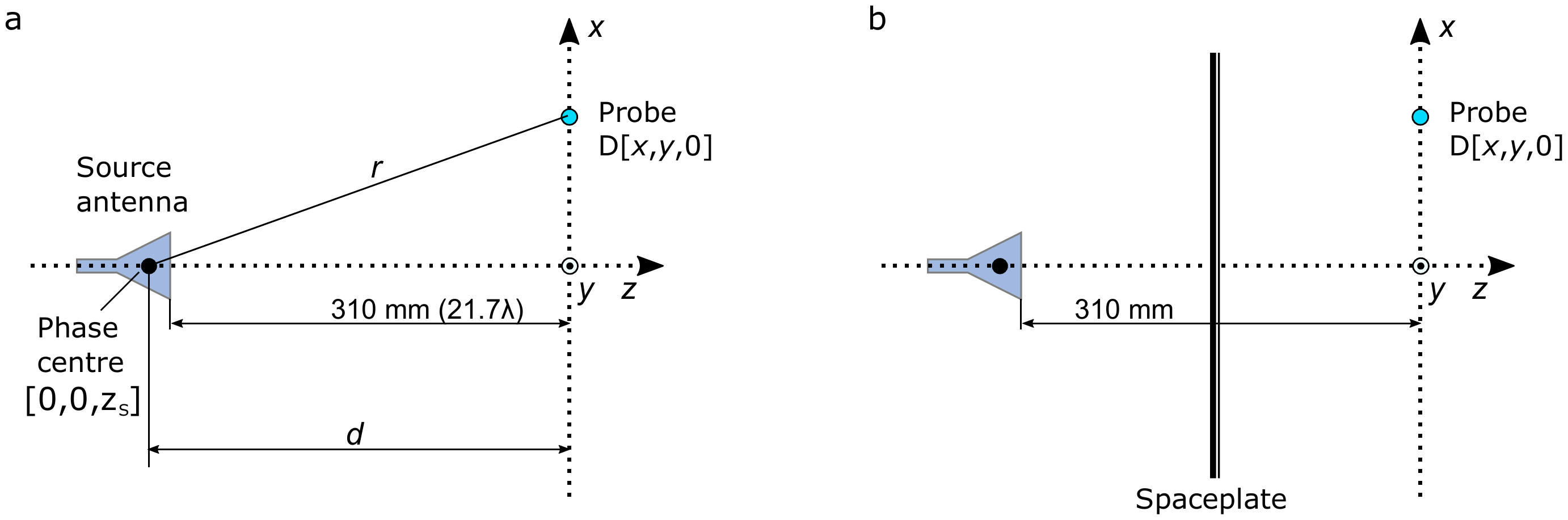}
\caption{Schematic view of the experimental setup for electric field scanning. Fig. a) represents the reference measurement, where the field radiated by the source antenna is measured on a plane by the detector (in fact we measure transmission coefficient S$_{21}$ which is proportional to the electric field). In b) the spaceplate is introduced in between the source antenna and the detection plane. The radiation from the source passing through the spaceplate is measured on two orthogonal cuts x = 0 and y = 0. All the dimensions are in millimeters and also relative to the free space wavelength at 21~GHz ($\lambda = 14.3$~mm).}
\label{fig:exp_setup_schematic}
\end{figure}

\noindent{\bf 4.3 Phase centre of an antenna and its determination}\\
The phase centre (PC) of an antenna is the key concept in our experimental demonstration of a spaceplate and is thus described in more detail here. The phase centre can be defined as an imaginary point associated with an antenna which appears as the origin of spherical wavefronts emanating from the antenna.  We present raw measured data as well as extracted positions of phase centres in a setup with and without the spaceplate - the difference is directly related to the compression factor.\\

Here, we briefly review how the phase centre of an antenna can be determined from the measured phase profile $\phi_{\mathrm{e}}$ (i.e. $ \mathrm{arg}\{S_{21}\}$). We demonstrate the process on phase measured along the x-axis $\phi_{\mathrm{e}}(x)$ which corresponds to the TM polarisation cut.
The task can be simply defined as a search for an optimum distance $d$ between a theoretical point source  and the centre of the xy-plane where the field is sampled that minimizes the root mean square error between the theoretical $\phi_{\mathrm{t}}(x)$ and experimental phase profiles $\phi_{\mathrm{e}}(x)$.

\begin{equation}
	\min_{d} \sqrt{ \frac{1}{N_x} \sum_{x_i=1}^{Nx} \left[ (\phi_{\mathrm{e}} - \phi_{\mathrm{t}})  - \mathrm{mean} (\phi_{\mathrm{e}} - \phi_{\mathrm{t}})  \right]^2  },
\end{equation}
where $N_x$ is the number of measurement points, $x_i$ are their locations and the distance $d$ is a function of angular frequency $d = d(\omega)$ as a result of frequency dependent properties of both the microwave antenna as well as of the spaceplate. The theoretical phase as a function of position $x$ is given as
\begin{equation}
	\phi_{\mathrm{t}}(x) = \mathrm{arg}\{ E(x) \} =  \mathrm{arg} \left\{ \frac{1}{r(x)} \exp (\mathrm{i}kr(x)) \right\}  = k r(x) = k \sqrt{d^2+x^2},
\end{equation}
where $E(x)$ is the electric field distribution along the x-direction, $r = |\overrightarrow{r}|$ is the distance between the phase centre and a point on the detection plane and $k = 2\pi/\lambda$ is the free space wave number (see Fig.~\ref{fig:exp_setup_schematic}).\\


\noindent{\bf 4.4 Results of E-field spatial mapping}\\
We present here the measured transmission phase $\phi_e$ and normalised magnitude $|t|/|t_{\mathrm{max}}|$ which are proportional to the magnitude and phase of the electric field on the measurement plane in a setup with and without the spaceplate (see Fig. \ref{fig:exp_setup_schematic}). The phase patterns show the unwrapped phase as a function of transverse position $x$ or $y$ for the measurement without the spaceplate (black curves) and with the spaceplate (red/blue curves). The measurements shown in Fig.~\ref{fig:measured_phase_and_magnitude}  correspond to two orthogonal planes where the polarisation of the wave is either TM (red) or TE (blue) with respect to the surface of the spaceplate, respectively. As a result of the apparent shift of the phase centre away from the measurement plane, we measure a shallower phase profile when the SP is introduced and a broader beam.  

\begin{figure}[ht]
\centering
\includegraphics[width=0.8\linewidth]{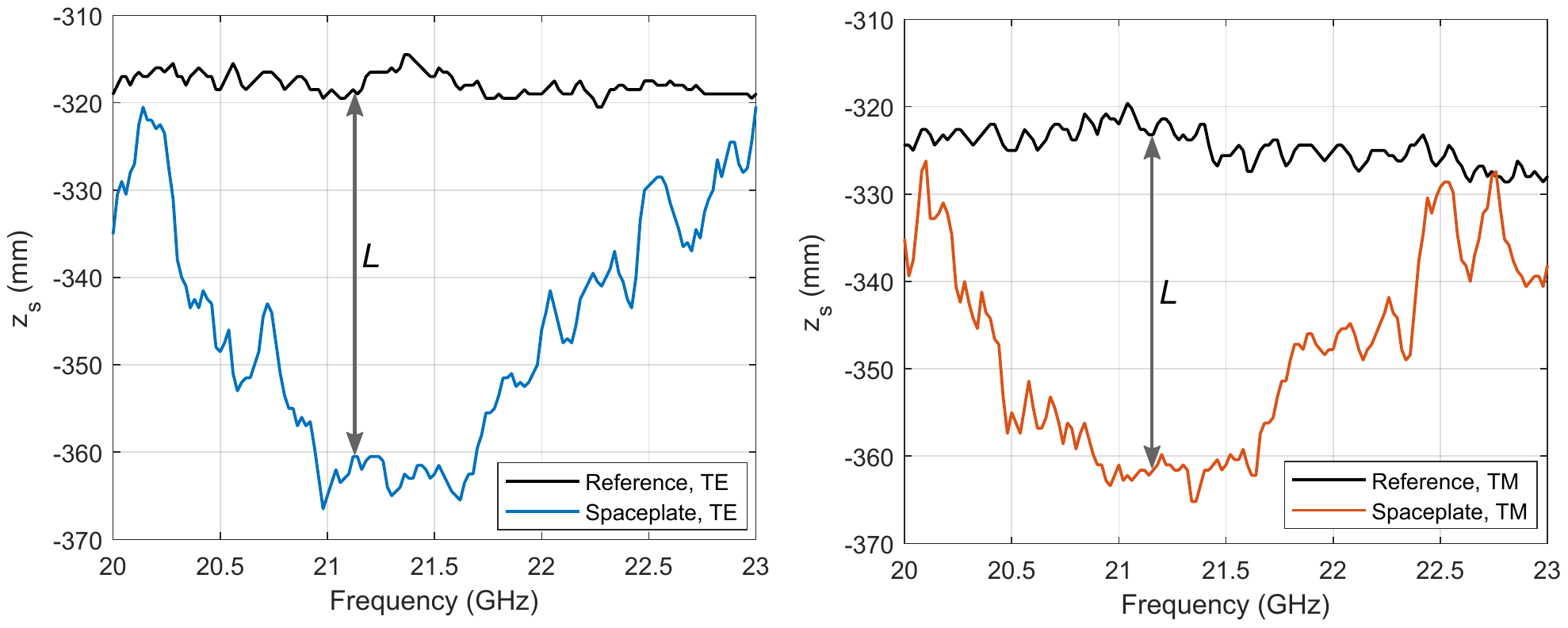}
\caption{Phase centre position as a function of frequency for TE (left) and TM (right) polarisation. The apparent added path length $L$ is introduced by the spaceplate (see Fig.~1 of the main paper). The compression factor corresponding to the length $L$ is given in Fig.~5 of the main paper. }
\label{fig:phase_centre_shift}
\end{figure}

\begin{figure}[ht]
\centering
\includegraphics[width=1\linewidth]{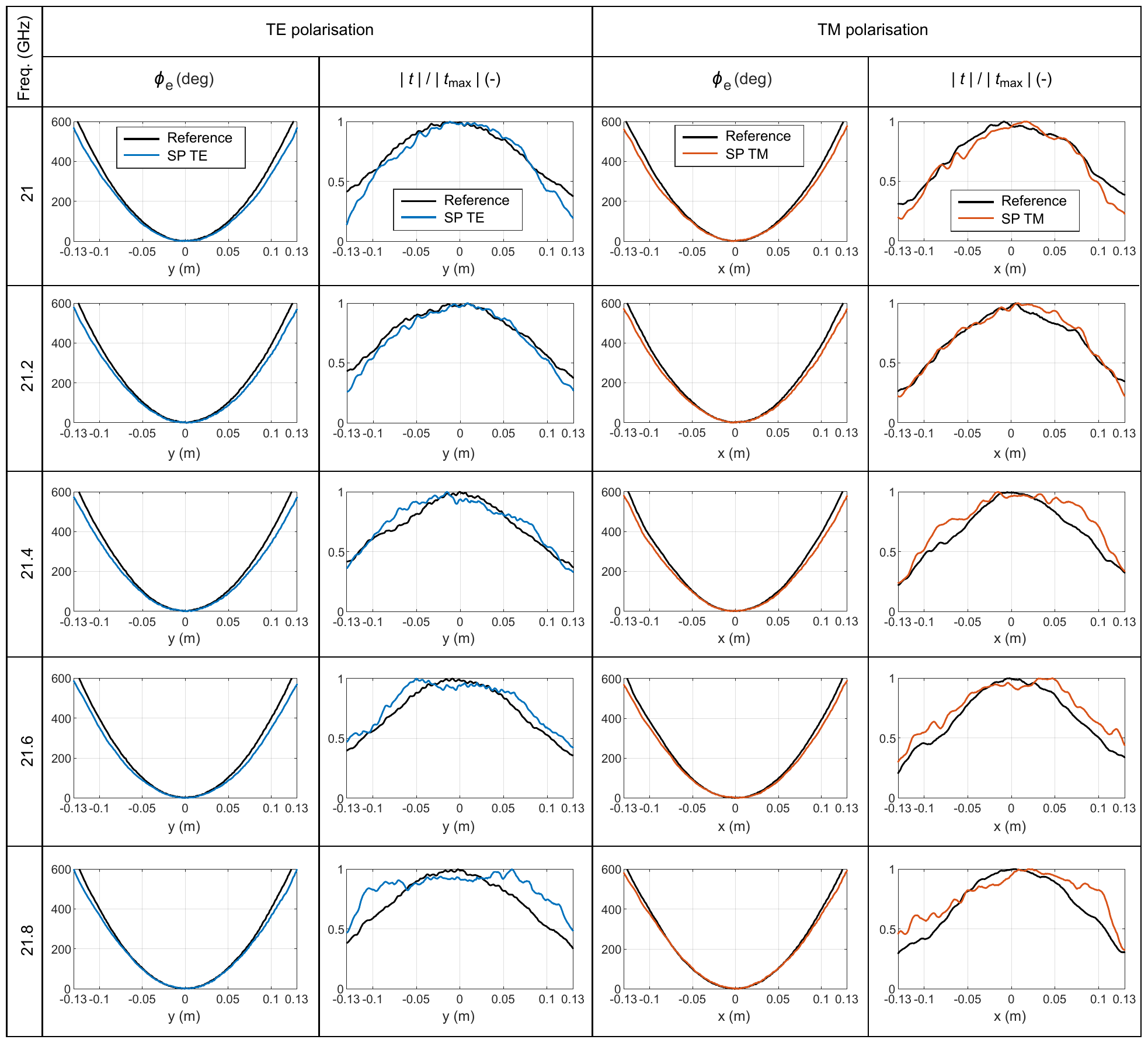}
\caption{Phase $\phi_e$  and normalized magnitude $|t|/|t_{\mathrm{max}}|$ of measured transmission coefficient (i.e. $t \equiv S_{21}$) between an illuminating antenna and an electrically small probe on a 1D grid according to Fig.~\ref{fig:exp_setup_schematic}. The coefficient  $t$ thus corresponds to E-field on two orthogonal cuts, for TE and TM polarisation. The effect of the spaceplate on the detected phase is clear, the shallower phase profile is a direct result of spherical wavefronts with larger radius of curvature compared to the measurement without the spaceplate (black curves).}
\label{fig:measured_phase_and_magnitude}
\end{figure}

The plots in Fig. \ref{fig:phase_centre_shift} include the information about the apparent position of the phase centre $z_\mathrm{S}$ calculated from the measured phase profiles using the method described in the previous section across the frequency range of 20 to 23~GHz. A more negative coordinate $z_\mathrm{S}$ corresponds to the source position apparently being farther away from the measurement plane.\\

\newpage
\noindent{\large \bf \S 5 Evolutionary optimization of a stochastic spaceplate}\\
\label{sec:stochastic}
A non-local metamaterial spaceplate can also be realised as a multilayer stack of homogeneous and isotropic layers distributed along the optical axis \cite{reshef2021}. The parameters of individual layers -- the thicknesses and refractive indices -- are then optimised by a stochastic optimisation algorithm. Such a design strategy does not inherently rely on any knowledge about the behavior of the elements. Theoretically (if the search space allows it), it can take advantage of  e.g.\ higher order modes within layers, creating multiple resonant cavities coupled by arbitrary coefficients (unlike the design in [3]) etc. However, the results of this approach presented in~\cite{reshef2021} tell us that finding a feasible solution can be quite challenging (for example in~\cite{reshef2021}, with 21 layers and $\mathcal{C}$ = 4.9, a spaceplate with an angular range of about 12-15 deg showed transmittance below $\sim$1.5\% for normal incidence). The other disadvantage is the non-scalability of the solution. Designing the spaceplate for other effective thicknesses requires, in general, starting the optimization from a scratch. A suitable optimisation strategy could be the combination of the global search with a local gradient-based optimization method.

The problem can be formulated as either a single- or multi-objective. In our experience, the single-objective formulation with three discernable sub-objectives (amplitude, phase, compression factor $\mathcal{C}$) led to a feasible solution more quickly. Generally, the objective is defined as the minimisation of the error between the desired transmission coefficient of the free space $t_0$ and a transmission coefficient of a spaceplate $t_{\mathrm{SP}}$, while the compression factor $\mathcal{C}$ should be maximized.  If we break the transmission coefficient into the amplitude and the phase derivative criterion and we sum through a finite number of incidence angles we can write for both polarisations (superscripts TE, TM)

\begin{equation*}
\begin{split}
 OBJ = \frac{1}{N}\sum_{\theta_i}^{N} c_1\cdot(|t^{\mathrm{TM}}_{\mathrm{SP}}(\theta_i)| - |t_0(\theta_i)|)^2 + c_2\left(\frac{\mathrm{d} arg(t^{\mathrm{TM}}_{\mathrm{SP}}(\theta_i))}{\mathrm{d} \theta}  - \frac{\mathrm{d} arg(t_{0}(\theta_i))}{\mathrm{d} \theta} \right)^2 + \\
\frac{1}{N}\sum_{\theta_i}^{N} c_3\cdot(|t^{\mathrm{TE}}_{\mathrm{SP}}(\theta_i)| - |t_0(\theta_i)|)^2 + c_4\left(\frac{\mathrm{d} arg(t^{\mathrm{TE}}_{\mathrm{SP}}(\theta_i))}{\mathrm{d} \theta}  - \frac{\mathrm{d} arg(t_{0}(\theta_i))}{\mathrm{d} \theta} \right)^2 + \\
c5\cdot d_{\mathrm{SP}} / d_{\mathrm{eff}}
\end{split}
\end{equation*}

Here, the derivatives are used to remove any constraints on the global phase of the transmitted field, which is irrelevant for most of the spaceplate applications. The five weighing coefficients $c_1$ - $c_5$ allow us to tune the importance of the amplitude ($c_1$, $c_3$) and phase ($c_2$, $c_4$) errors and the maximum achieved compression factor (1/$c_5$). The inputs of the algorithm are the maximum number or the layers and the effective thickness $d_\mathrm{eff}$ which we are trying to squeeze. During the optimisation, the transmission coefficient of the free-space is pre-calculated (as it does not change during the optimisation) and the transmission coefficient of the SP is evaluated by a transfer (characteristic) matrix method \cite{hecht}, which gives a full-wave field solution for a multilayer stack.

In our design, we tried to reduce the search space as much as possible while still producing a solution with relatively high $\mathcal{C}$ (in this case we set  $\mathcal{C}>3$). With the experimental demonstration in mind, we opted for a two material combination with alternating high/low refractive index medium – as the low $n$ medium, we selected air and for the high $n$ medium a commercially available microwave substrate from Rogers ($\varepsilon_r$ = 10.2). We fixed the thicknesses of the high $n$ dielectric layers to 1.52\,mm (thickness of the substrate). Thus, the optimization parameters are only the thicknesses of the air gaps and the total number of the layers. By allowing the thicknesses of the layers to go to zero (skipping the layer), the optimiser can effectively double the maximum thickness limitations defined during the initialisation. 

In the optimisation results below, we set the frequency of operation to $f$ = 15\,GHz,  maximum number of layers ($NL_{\mathrm{max}}$) to 15, the effective thickness to be substituted by SP to $d_\mathrm{eff}$ = 250\,mm. An evolutionary optimization strategy based on a genetic algorithm in Matlab was used to search for feasible solutions. With maximum of 15 layers (8 layers of microwave substrate of fixed thickness, separated by 7 layers of air), we are optimizing only 7 parameters at most.  Thus, the population does not have to be very large - we usually worked with about 100 individuals per population. On a laptop with Intel i7-7500, 2.7\,GHz processor the evaluation of the forward model took on average 5\,ms for both polarisations, 15 layers and a single frequency point. Evaluating the whole population thus took only about 0.5\,s.

\begin{figure}[ht]
\centering
\includegraphics[width=0.5\linewidth]{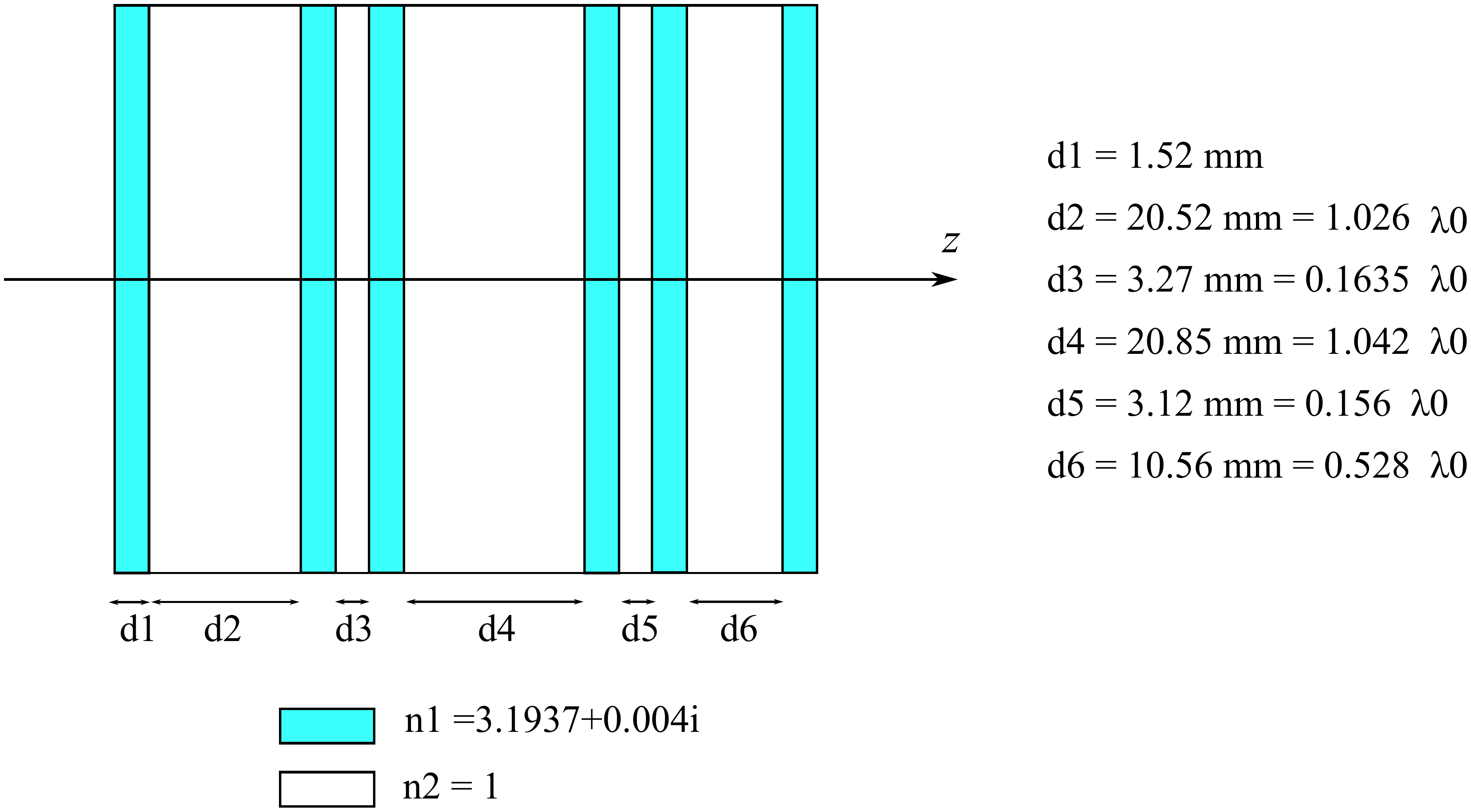}
\caption{Spaceplate designed by a genetic algorithm with total thickness of $d_\mathrm{SP}=$ 67.44 mm substitutes a slab of air with equivalent thickness $d_\mathrm{eff} = 250$ mm corresponding to a compression factor $\mathcal{C} \approx 3.7$.}
\label{fig:multilayer_stack}
\end{figure}

One of the optimised structures is shown in Fig. \ref{fig:multilayer_stack}. We can see that the first, third and fifth gaps have lengths very close to the even FP resonances whereas the second and the fourth are relatively small. The resemblance of this stochastically optimised structure to the empirical multi-cavity design proposed in \cite{chen2021} is clear.

\begin{figure}[ht!]
\centering
\includegraphics[width=0.75\linewidth]{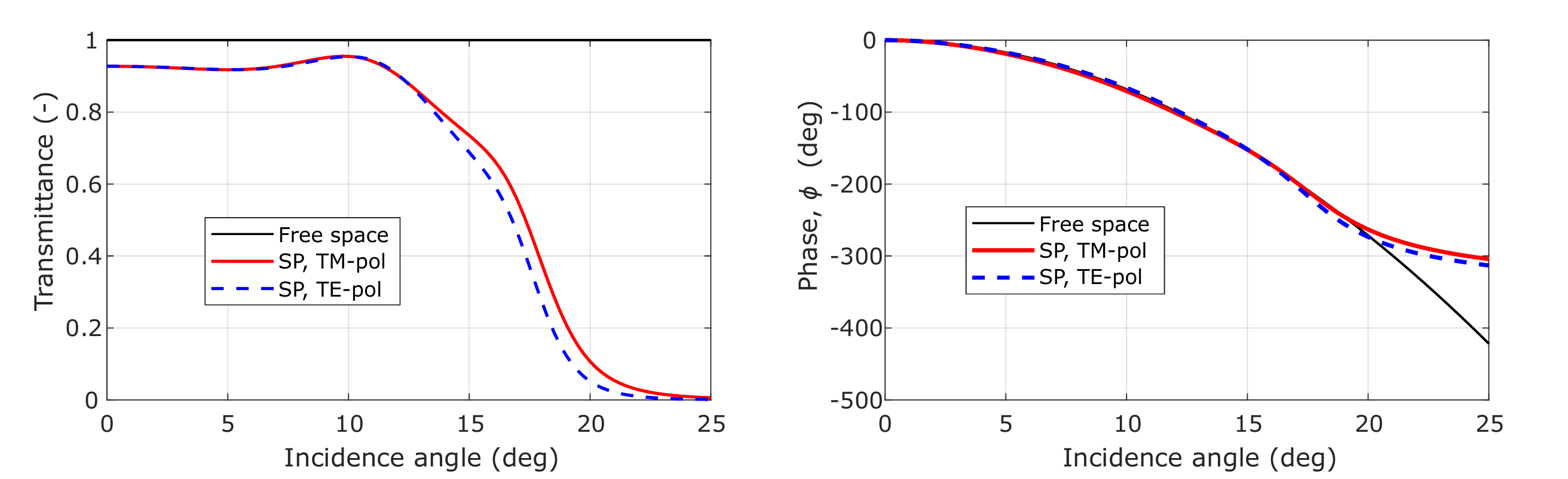}
\caption{Transmittance and the transmission phase of the spaceplate as a function of incidence angle at 15 GHz. The free space fit corresponds to a distance $d_\mathrm{eff} = 250$ mm.}
\label{fig:stoc_results_1D}
\end{figure}

\begin{figure}[ht!]
\centering
\includegraphics[width=0.78\linewidth]{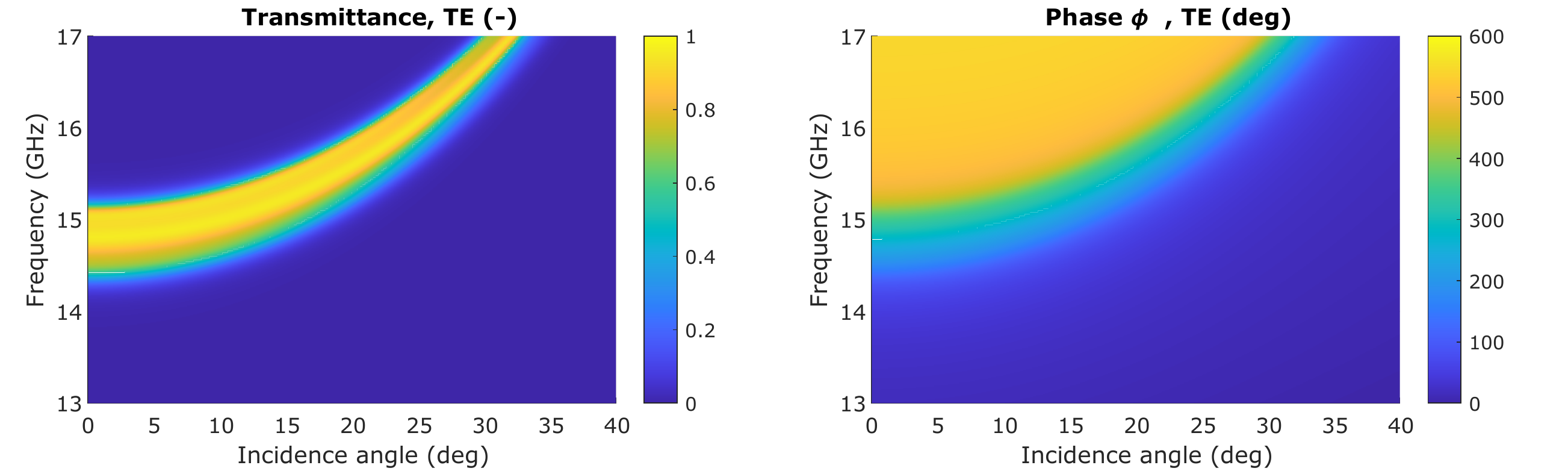}
\caption{Transmittance and the phase of the spaceplate as a function of incidence angle and frequency for TE polarisation.}
\label{fig:stoc_results_2Da}
\end{figure}

\begin{figure}[ht!]
\centering
\includegraphics[width=0.78\linewidth]{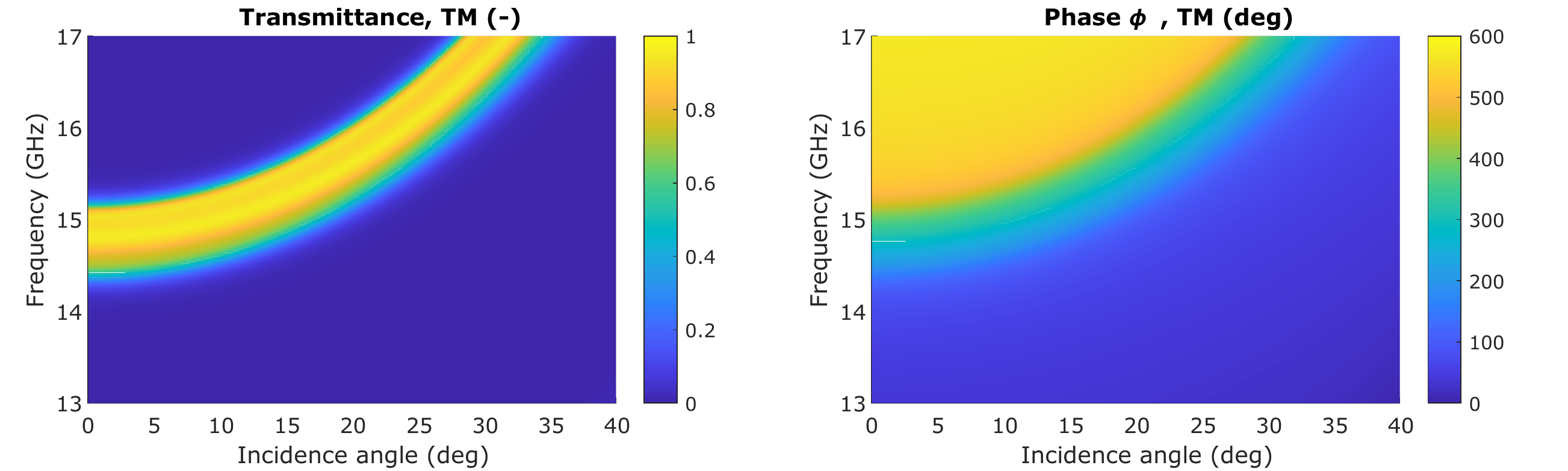}
\caption{Transmittance and the phase of the spaceplate as a function of incidence angle and frequency for TM polarisation.}
\label{fig:stoc_results_2Db}
\end{figure}

Figure \ref{fig:stoc_results_1D} gives the transmittance $|t|^2$ and the phase of the transmission coefficient for the two polarisations and compares them to the free-space fit. In Fig. \ref{fig:stoc_results_2Da} and \ref{fig:stoc_results_2Db} we can see the transmittance and the transmission phase as functions of the frequency and the incidence angle.

\end{document}